\documentclass[preprint,aps,showpacs,amsmath,amssymb,prd,eqsecnum]{revtex4-1}
\usepackage{color}

\usepackage{natbib}
\allowdisplaybreaks

\newcommand{\nablaslash}{\nabla{\!\!\!\!\slash}}

\newcommand{\pslash}{p{\hspace{-5pt}\slash}}
\newcommand{\qslash}{q{\hspace{-5pt}\slash}}
\newcommand{\bare}{\textrm{B}}

\begin{document}
\bibliographystyle{revtex}

\title{Gauged Yukawa model in curved spacetime}

\author{David J. Toms}
\homepage{http://www.staff.ncl.ac.uk/d.j.toms}
\email{david.toms@newcastle.ac.uk}
\affiliation{
School of Mathematics, Statistics and Physics,
Newcastle University,
Newcastle upon Tyne, U.K. NE1 7RU}

\date{\today}

\begin{abstract}
The Yukawa model in curved spacetime is considered. We consider a complex scalar field coupled to a $U(1)$ gauge field and also interacting with Dirac fields with a general Yukawa coupling. The local momentum space method is used to obtain the one-loop effective action and we adopt the gauge condition independent background field method introduced by Vilkovisky and DeWitt. The pole parts of the one-loop effective action that depend on the background scalar field, that we do not assume to be constant, are found and used to calculate the counterterms and to determine the relevant renormalization group functions. Terms in the effective action that involve the gradient terms in the scalar field as well as the effective potential are found in the case where the scalar field and Dirac fields are massless. We also discuss the anomaly that arises if the pseudoscalar mass term for the Dirac fermions is removed by a chiral transformation.
\end{abstract}


\maketitle

\section{Introduction}\label{sec-intro}

The study of a general Yukawa model in curved spacetime was given recently~\cite{Tomsyukawa1}. The principle aim of the present paper is to generalize this analysis by the inclusion of a gauge field. The issue of the possible dependence on the gauge condition is addressed through the use of the Vilkovisky-DeWitt \cite{Vilkovisky1,DeWitt6} formalism. We obtain a result for the effective action, including terms that involve the background scalar field gradient, at one-loop order up to and including terms that are quadratic in the curvature. This means that we do not restrict the background scalar field to be constant.

There is some previous work on Yukawa interactions with scalars in curved spacetime. A selected set of references includes \cite{shapiro1989asymptotic,odintsov1992general,elizalde1994renormalization,elizalde1995higgs, elizalde1995improved,elizalde1995renormalization,ProkopecWoodard,Garbrechtfermion,Garbrecht,Miao,Shapiro2011PLB, herranen2014spacetime,Cz,markkanen20181}. In particular, \cite{markkanen20181} has looked at the renormalization group improved effective potential for the standard model in some detail and shown the potential importance of the $R^2$ terms in cosmology.  The generalization of the original Coleman and Weinberg~\cite{ColemanandWeinberg} analysis to curved spacetime was originally given by \cite{BuchbinderandOdintsovRG} but with the neglect of the $R^2$ terms. (See the overview in \cite{ParkerTomsbook} based on earlier analysis in Ref.~\cite{TomsRG}.) Yukawa interactions have also been considered in the asymptotic safety program for quantum gravity~\cite{Zanusso,Eichhornetal1,Eichhornetal3,oda2016non,Eichhornetal2}, as well as in perturbative quantum gravity~\cite{RodigastSchuster} including unimodular gravity~\cite{Martins2017PLB,Martins2018JCAP,Martins2018EPJC} and scale-invariant gravity~\cite{narain2017exorcising}.  

The outline of our paper is as follows. The general model of a charged scalar field interacting with a gauge filed and a spinor field through a Yukawa interaction is given in Sec.~\ref{sec2}. Both scalar and pseudoscalar mass terms are included, as are scalar and pseudoscalar Yukawa couplings. A brief description of the background field method~\cite{DeWittdynamical,Vilkovisky1,DeWitt6,ParkerTomsbook} is given in Sec.~\ref{secBose}, and the formal expression for the one-loop effective action is given. All the pole parts for the one-loop effective action coming from the vector and scalar fields are found. The local momentum space method originated by Bunch and Parker~\cite{BunchParker} is used. In Sec.~\ref{secFermi} we evaluate the pole part of the one-loop effective action that arises from the spinor fields in two ways: one method uses a perturbative approach like that for the Bose fields in Sec.~\ref{secBose}; the other uses a direct functional integral evaluation. Both methods are shown to agree. In Sec.~\ref{ctrg} we combine all the results for the pole terms from previous sections and work out the necessary counterterms and renormalization group functions. The renormalization group is used to evaluate the effective potential and the gradient part of the effective action in the case where neither the scalar field nor the spinor field have mass. We present a short discussion in the last section and comment on an anomaly that arises~\cite{Tomsyukawa1} if the pseudoscalar mass term for the fermions is removed by a chiral transformation. Some of the technical calculations are contained in the Appendices.

\section{The Gauged Yukawa model in curved spacetime}\label{sec2}

We will consider a complex scalar field $\Phi(x)$ coupled to a $U(1)$ gauge field $A_\mu(x)$ in a gauge invariant way. To have a gauge invariant Yukawa interaction we must consider two Dirac fields, one uncharged that we will call $\chi(x)$, and one charged that we will call $\Psi(x)$. We will allow the possibility of both a scalar and a pseudoscalar Yukawa couplings. The spacetime dimension will be four and we will adopt dimensional regularization~\cite{tHooftandVeltman}. All our conventions for curvature and spinors will follow those in Parker and Toms~\cite{ParkerTomsbook}. The bare action will be taken to be (with the subscript `B' signifying a bare quantity)
\begin{equation}
S=S_{\textrm {scalar}}+S_{\textrm{vector}}+S_{\textrm{spinor}}+S_{\textrm{grav}},\label{2.1}
\end{equation}
where
\begin{subequations}\label{2.2}
\begin{align}
S_{\textrm{scalar}}&=\int dv_x \Big\lbrack (D_\mu\Phi_{\textrm{B}})^\dagger(D^\mu\Phi_{\textrm{B}}) -m_{\textrm{s\,B}}^2\, |\Phi_{\textrm{B}}|^2 -\xi_{\textrm{B}}\,R \,|\Phi_{\textrm{B}}|^2 -\frac{\lambda_{\textrm{B}}}{6}\,|\Phi_{\textrm{B}}|^4\Big\rbrack,\label{2.2a}\\
S_{\textrm{vector}}&=-\frac{1}{4}\int dv_x \,F^{\mu\nu}F_{\mu\nu},\label{2.2b}\\
S_{\textrm{spinor}}&=\int dv_x\Big\lbrack \bar{\chi}(i\gamma^\mu\nabla_\mu-m_{\chi}-im_{\chi5}\gamma_5)\chi+\bar{\Psi}(i\gamma^\mu D_\mu-m_{\psi}-im_{\psi5}\gamma_5)\Psi\nonumber\\
&\qquad\qquad-\Phi^\dagger\bar{\chi}(w+iw_5\gamma_5)\Psi-\Phi\bar{\Psi}(w^\ast+iw_5^\ast\gamma_5)\chi\Big\rbrack,\label{2.2c}\\
S_{\textrm{grav}}&=\int dv_x\Big(\Lambda_{\textrm{B}}+\kappa_{\textrm{B}}\,R+\alpha_{1\,\textrm{B}}\,R^{\mu\nu\lambda\sigma}R_{\mu\nu\lambda\sigma} + \alpha_{2\,\textrm{B}}\,R^{\mu\nu}R_{\mu\nu} + \alpha_{3\,\textrm{B}}\,R^2 \Big).\label{2.2d}
\end{align}
\end{subequations}
We use $dv_x$ to stand for the invariant spacetime volume element: $dv_x=|\det g_{\mu\nu}(x)|^{1/2}d^nx$. The gauge covariant derivative $D_\mu$ is defined by 
\begin{equation}
D_\mu=\nabla_\mu-ieA_\mu,\label{2.3}
\end{equation}
where $\nabla_\mu$ is the spacetime covariant derivative. The gauge and spinor fields, as well as the spinor mass terms and Yukawa couplings are also bare but as we will not need to consider their renormalization here we will not indicate this explicitly. The theory is invariant under the local $U(1)$ gauge transformation
\begin{subequations}\label{2.4}
\begin{align}
\Phi(x)&\rightarrow e^{ie\theta(x)}\,\Phi(x),\label{2.4a}\\
\Psi(x)&\rightarrow e^{ie\theta(x)}\,\Psi(x),\label{2.4b}\\
A_\mu(x)&\rightarrow A_\mu(x)+\nabla_\mu\,\theta(x),\label{2.4c}\\
\chi(x)&\rightarrow\chi(x).\label{2.4d}
\end{align}
\end{subequations}
$F_{\mu\nu}=\nabla_\mu A_\nu -\nabla_\nu A_\mu$ is the usual field strength tensor. The Yukawa coupling constants are denoted by $w$ and $w_5$ and can be taken as arbitrary complex numbers. We allow for the possibility of both scalar and pseudoscalar couplings. Likewise, we include both scalar and pseudoscalar mass terms for the spinor fields. As discussed in \cite{Tomsyukawa1} it is possible to remove the pseudoscalar mass term by a chiral transformation on the spinor fields; however, the effective action is not invariant under this change due to an anomaly. This anomaly does not affect the one-loop counterterms but we will leave the action in the form given above. $\gamma^\mu$ are the spacetime dependent Dirac matrices~\cite{ParkerTomsbook} defined in terms of the usual Minkowski ones~\cite{BjorkenandDrell} $\gamma^a$ by $\gamma^\mu=e_{a}{}^{\mu} \gamma^a$ with $e_{a}{}^{\mu}$ the vierbein. (Latin letters will denote orthonormal frame indices.) $\gamma_5$ is Hermitian with $\gamma_5^2=I$ where $I$ the identity matrix, and constant (since it is defined in terms of the local orthonormal frame $\gamma$-matrices). The factors of $i$ in \eqref{2.2c} ensure that the action is real. The gravitational part of the action, $S_{\textrm{grav}}$, is required to deal with the vacuum part of the effective action~\cite{BirrellandDavies,ParkerTomsbook}.

Our aim here is to calculate the curved spacetime effective potential to one-loop order using the renormalization group. We will therefore expand about a general scalar field background but set the background gauge and spinor fields to zero. Because we are only working to one-loop order we only need to keep terms in the expansion of the action that are quadratic in the quantum fields (which are those that are integrated over in the functional integral that defines the effective action). It is simplest to do this if we write the complex scalar field in terms of its real and imaginary parts as
\begin{equation}
\Phi=\frac{1}{\sqrt{2}}\,(\Phi_1+i\,\Phi_2).\label{B1}
\end{equation}
The $U(1)$ gauge symmetry now becomes an $O(2)$ symmetry for $(\Phi_1,\Phi_2)$. We can use this symmetry to take the background scalar field to lie in the $\Phi_1$ direction without any loss of generality. We will therefore take the background field expansion of the complex scalar field $\Phi$ to be
\begin{equation}
\Phi=\frac{1}{\sqrt{2}}\,(\varphi+\phi_1+i\,\phi_2),\label{B2}
\end{equation}
where $\varphi(x)$ is the background scalar field, which is real, and $\phi_1,\phi_2$ are the quantum fields that are integrated over in the functional integral. 

Using \eqref{B2} we find from \eqref{2.2} that the terms in the action that are quadratic in the quantum fields are
\begin{align}
S_{\textrm{quad}}&=\frac{1}{2}\,\int dv_x\Big\lbrace A_\mu\Box A^\mu+R^{\mu\nu}A_\mu A_\nu +(\nabla^\mu A_\mu)^2-\phi_1(\Box+m_s^2+\xi R)\phi_1\nonumber\\
&\quad-\phi_2(\Box+m_s^2+\xi R)\phi_2-\frac{\lambda}{4}\,\varphi^2\phi_1^2-\frac{\lambda}{12}\,\varphi^2\phi_2^2+e\,A^\mu\phi_2\nabla_\mu\varphi\nonumber\\
&\quad-e\,A^\mu\varphi\nabla_\mu\phi_2+\frac{e^2}{2}\,\varphi^2 A^\mu A_\mu\Big\rbrace\nonumber\\
&+\int dv_x\Big\lbrace \bar{\chi}(i\gamma^\mu\nabla_\mu-m_{\chi}-im_{\chi5}\gamma_5)\chi+\bar{\Psi}(i\gamma^\mu \nabla_\mu-m_{\psi}-im_{\psi5}\gamma_5)\Psi\nonumber\\
&\qquad\qquad-\frac{1}{\sqrt{2}}\varphi\bar{\chi}(w+iw_5\gamma_5)\Psi-\frac{1}{\sqrt{2}}\varphi\bar{\Psi}(w^\ast+iw_5^\ast\gamma_5)\chi\Big\rbrace.\label{B3}
\end{align}
Because there are no direct couplings between the quantum Fermi fields $(\chi,\Psi)$ and the quantum Bose fields $(A_\mu,\phi_1,\phi_2)$ the functional integral will factorize. The contributions from the Bose and Fermi fields can be considered separately and this will be done in Secs.~\ref{secBose} and \ref{secFermi} respectively.

\section{Contribution to the effective action from Bose fields}\label{secBose}

To compute the contribution of Bose fields to the effective action we must first choose a gauge condition. We will utilize the Vilkovisky-DeWitt method~\cite{Vilkovisky1,DeWitt6} here. All our conventions and notation will follow \cite{ParkerTomsbook} where a more detailed description can be found. The formalism gives a result for the effective action that is completely independent of the choice of gauge condition. Although it is possible to proceed with a general gauge choice, as emphasized originally by Fradkin and Tseytlin~\cite{FradkinTseytlin} calculations are considerably simpler if a special gauge condition is chosen, namely the Landau-DeWitt gauge. If we let $\varphi^i$ represent the complete set of fields in condensed notation, then the set of gauge transformations can be written as
\begin{equation}
\delta\varphi^i=K^{i}_{\alpha}\lbrack\varphi\rbrack\delta\epsilon^\alpha,\label{B4}
\end{equation}
for some $K^{i}_{\alpha}$. Here $\delta\epsilon^\alpha$ represent the infinitesimal parameters of the gauge transformation. From \eqref{2.4} we have (using \eqref{B1})
\begin{subequations}\label{B5}
\begin{align}
\delta\phi_1&=-e\delta\epsilon\phi_2,\label{B5a}\\
\delta\phi_2&=e\delta\epsilon\phi_1,\label{B5b}\\
\delta A_\mu&=\nabla_\mu\delta\epsilon.\label{B5c}
\end{align}
\end{subequations}
The expressions for $K^{i}_{\alpha}$ can be simply read off from comparison with \eqref{B4}.

The central idea behind the Vilkovisky-DeWitt method is to consider a metric on the space of fields and use this to construct a connection. The effective action can then be obtained in a completely covariant way that is independent of how the fields are parametrized, as well as independent of the gauge condition. The field space metric can be obtained from the derivative terms in the action by analogy with the nonlinear sigma model (where a covariant approach is both obvious and natural). Because of the sign difference in the $\Box$ terms of \eqref{B3} for the gauge and scalar fields we can choose the field space metric, $g_{ij}$ in condensed notation, to be
\begin{equation}
g_{ij}=\left(\begin{array}{cc}-g^{\mu\nu}&0\\ 0&I\\\end{array}\right)\delta(x,x'),\label{B6}
\end{equation}
with the choice $(A_\mu,\phi_1,\phi_2)$ for the condensed index expression $\varphi^i$. With our conventions, if we perform a Wick rotation to imaginary time $g_{ij}$ becomes positive definite, hence the overall sign choice in \eqref{B6}.

If we now write the background field expansion generally as
\begin{equation}
\varphi^i=\varphi^i_\star+\eta^i,\label{B7}
\end{equation}
where $\varphi^i_\star$ is the background field and $\eta^i$ is the quantum field, then the Landau-DeWitt gauge condition reads
\begin{equation}
g_{ij}\lbrack\varphi_\star\rbrack\,K^{i}_{\alpha}\lbrack\varphi_\star\rbrack\,\eta^j=0.\label{B8}
\end{equation}
If we uncondense the indices in \eqref{B8} we have
\begin{equation}
\nabla^\mu A_\mu+e\,\varphi\,\phi_2=0,\label{B9}
\end{equation}
for the case under consideration here. The gauge condition can be enforced with a $\delta$-function in the functional integral that defines the effective action along with a Faddeev-Popov factor that follows from \eqref{B9} in the usual way by considering the change in \eqref{B9} under the infinitesimal gauge transformation \eqref{B5}. The background field is held fixed for this. The resulting Faddeev-Popov determinant is $\det(\Box+e^2\,\varphi^2)$.

The Vilkovisky-DeWitt connection has two main terms. The first is the Christoffel connection that follows from the field space metric $g_{ij}$. Because there is no dependence on the fields in \eqref{B6} this will vanish in our case. The second term in the connection involves the gauge transformation generator $K^i_\alpha$ and its derivative. (The exact expression can be found in \cite[page 378]{ParkerTomsbook} for example.) However as noted by Fradkin and Tseytlin~\cite{FradkinTseytlin} this term makes no contribution to the effective action at one-loop order if we adopt the Landau-DeWitt gauge condition. (This is very easy to see at one-loop order and an inductive proof~\cite[Sec.~7.5.1]{ParkerTomsbook} shows that it holds to all orders in the loop expansion.) The net result is that we may now proceed as usual in our evaluation of the effective action with the assurance that everything is covariant and independent of the gauge condition. If any other gauge choice than Landau-DeWitt is made then the calculation is more involved as the full Vilkovisky-DeWitt connection must be used.

The one-loop effective action coming from the Bose fields $\Gamma^{(1)}_{\textrm{Bose}}$ is given by
\begin{equation}
e^{i\,\Gamma^{(1)}_{\textrm{Bose}}}=\int\lbrack d\varphi^i\rbrack\,\delta\lbrack \nabla^\mu A_\mu+e\,\varphi\,\phi_2\rbrack\,\det(\Box+e^2\,\varphi^2)\,e^{i\,S^{\textrm{Bose}}_{\textrm{quad}}},\label{B10}
\end{equation}
where $\lbrack d\varphi^i\rbrack$ denotes integration over the Bose fields $(A_\mu,\phi_1,\phi_2)$, and $S^{\textrm{Bose}}_{\textrm{quad}}$ is the quadratic part of \eqref{B3} coming from just the Bose fields. The $\delta$-function in the integrand can be exponentiated using the standard representation for the $\delta$-function
\begin{equation}
\delta(x)=\lim_{\alpha\rightarrow0}(-2\pi i\alpha)^{-1/2}e^{-\frac{i}{2\alpha}\,x^2},\label{B11}
\end{equation}
extended to functions. Ignoring any overall constants that can be absorbed into the functional measure we then have
\begin{equation}
e^{i\,\Gamma^{(1)}_{\textrm{Bose}}}=\det(\Box+e^2\,\varphi^2)\,\int\lbrack d\varphi^i\rbrack\,\,e^{i\,\tilde{S}^{\textrm{Bose}}_{\textrm{quad}}},\label{B12}
\end{equation}
where
\begin{equation}
\tilde{S}^{\textrm{Bose}}_{\textrm{quad}}={S}^{\textrm{Bose}}_{\textrm{quad}} - \frac{1}{2\alpha}\,\int dv_x (\nabla^\mu A_\mu+e\,\varphi\,\phi_2)^2.\label{B13}
\end{equation}
The $\alpha\rightarrow0$ limit is understood in \eqref{B12}.

We are interested in terms in $\Gamma^{(1)}_{\textrm{Bose}}$ that contain poles that involve the background scalar field $\varphi$. We can write
\begin{equation}
\tilde{S}^{\textrm{Bose}}_{\textrm{quad}}==S_0+S_1+S_2,\label{B14}
\end{equation}
where the subscript on the right-hand side counts the power of $\varphi$ that occurs. From \eqref{B3} and \eqref{B13} we have
\begin{subequations}\label{B15}
\begin{align}
S_0&=\frac{1}{2}\,\int dv_x\Big\lbrack A_\mu\Box A^\mu+R^{\mu\nu}A_\mu A_\nu -\Big(1-\frac{1}{\alpha}\Big)A_\mu\nabla^\mu\nabla^\nu A_\nu\nonumber\\
&\quad-\phi_1(\Box+m_s^2+\xi R)\phi_1 -\phi_2(\Box+m_s^2+\xi R)\phi_2\Big\rbrack,\label{B15a}\\
S_1&=\int dv_x\Big\lbrack e\,\Big(1+\frac{1}{\alpha}\Big)\,A_\mu\,\phi_2\,\nabla^{\mu}\varphi - e\,\Big(1-\frac{1}{\alpha}\Big)\,\varphi\,A_\mu\,\nabla^{\mu}\phi_2\Big\rbrack,\label{B15b}\\
S_2&=\int dv_x\Big\lbrack -\frac{\lambda}{4}\,\varphi^2\,\phi_1^2-\Big(\frac{\lambda}{12}+\frac{e^2}{2\alpha}\Big)\,\varphi^2\,\phi_2^2 + \frac{e^2}{2}\,\varphi^2\,A^\mu A_\mu\Big\rbrack.\label{B15c}
\end{align}
\end{subequations}
Because we will be calculating the effective potential using the renormalization group functions it is essential to consider any possible renormalization of the background scalar field $\varphi$. This means it is not allowed to assume that $\varphi$ is constant as this will miss any possible field renormalization.

The term in $S_0$ will contribute to the vacuum part of the effective action that we will consider later. We will initially concentrate on those terms in $\Gamma^{(1)}_{\textrm{Bose}}$ that involve the background scalar field $\varphi$ with the goal of identifying the renormalization counterterms and the scalar field renormalization factor so that the renormalization group functions can be found. Because the integrand in \eqref{B12} involves a Gaussian it is possible to use the heat kernel method but this is complicated by the fact that the resulting operator is not diagonal in the fields, and additionally that the operator for the vector fields is not minimal due to the presence of the $A_\mu\nabla^\mu\nabla^\nu A_\nu$ term in \eqref{B15a}. It would be possible to use the method of Barvinsky and Vilkovisky~\cite{BarvinskyVilkovisky} or else of Moss and Toms~\cite{MossToms} here, but we will instead make use of a different method that makes more contact with a traditional Feynman diagram analysis.

We will treat $S_1+S_2$ as the interaction part of the action with $S_0$ determining the Green's functions or propagators. The one-loop effective action in \eqref{B12} becomes
\begin{equation}
\Gamma^{(1)}_{\textrm{Bose}}=-i\,\ln\det(\Box+e^2\,\varphi^2)-i\,\langle e^{i\,(S_1+S_2)}\rangle,\label{B16}
\end{equation}
where $\langle\cdots\rangle$ means to evaluate using Wick's theorem with only terms corresponding to connected diagrams kept. The basic Green functions are defined by
\begin{subequations}\label{B17}
\begin{align}
\langle A_\mu(x)A_\nu(x')\rangle&=i\,G_{\mu\nu}(x,x'),\label{B17a}\\
\langle \phi_1(x)\phi_1(x')\rangle&=\langle \phi_2(x)\phi_2(x')\rangle=i\,\Delta(x,x'),\label{B17b}
\end{align}
\end{subequations}
where
\begin{subequations}\label{B18}
\begin{align}
\Big\lbrack g^{\mu\lambda}\Box+R^{\mu\lambda}-\Big(1-\frac{1}{\alpha}\Big)\nabla^\mu\nabla^\lambda\Big\rbrack\,G_{\lambda\nu}(x,x')&=\delta^{\mu}_{\nu}\,\delta(x,x'),\label{B18a}\\
(-\Box-m_s^2-\xi R)\,\Delta(x,x')&=\delta(x,x').\label{B18b}
\end{align}
\end{subequations}
Terms like $\langle A_{\mu}(x)\phi_1(x')\rangle$ that involve different Bose fields will vanish because $S_0$ is diagonal in the fields. When the exponential in \eqref{B16} is expanded in powers of $S_1+S_2$ this will ensure that all terms that are odd in $\varphi$ will vanish. By simple power counting, which is valid at one-loop order, all terms that involve $\varphi$ with a power that exceeds four will be finite and contain no pole terms in dimensional regularization; these terms can make no contribution to the renormalization group functions. If we denote the pole part of any expression by $\textrm{PP}\lbrace\cdots\rbrace$ it then follows from \eqref{B16} that
\begin{equation}
\textrm{PP}\left\lbrace \Gamma^{(1)}_{\textrm{Bose}}\right\rbrace=-i\,\textrm{PP}\left\lbrace\ln\det(\Box+e^2\,\varphi^2)\right\rbrace+ \textrm{PP}\left\lbrace \Gamma_2\right\rbrace+ \textrm{PP}\left\lbrace\Gamma_4 \right\rbrace,\label{B19}
\end{equation}
where
\begin{equation}
\Gamma_2=\langle S_2\rangle+\frac{i}{2}\langle S_1^2\rangle,\label{B20}
\end{equation}
is quadratic in $\varphi$, and
\begin{equation}
\Gamma_4=\frac{i}{2}\langle S_2^2\rangle-\frac{1}{2}\langle S_1^2S_2\rangle-\frac{i}{24}\langle S_1^4\rangle, \label{B21}
\end{equation}
is quartic in $\varphi$. We will evaluate the pole parts of $\Gamma_2$ and $\Gamma_4$ in the next two subsections. The first term on the right hand side of \eqref{B21} will be evaluated in Sec.~\ref{secghost}.

\subsection{$\textrm{PP}\left\lbrace \Gamma_2\right\rbrace$}\label{PPGamma2}

By using \eqref{B15b} along with \eqref{B17} it can be shown that
\begin{align}
\langle S_1^2\rangle&=-e^2\int dv_xdv_{x'}\Big\lbrack \Big(1+\frac{1}{\alpha}\Big)^2\nabla^\mu\varphi(x) \nabla^{\prime\nu}\varphi(x')\, G_{\mu\nu}(x,x')\Delta(x,x')\nonumber\\
&-2\Big(1-\frac{1}{\alpha^2}\Big)\varphi(x)\nabla^{\prime\nu}\varphi(x')\,G_{\mu\nu}(x,x')\nabla^\mu\Delta(x,x')\nonumber\\
&+\Big(1-\frac{1}{\alpha}\Big)^2\varphi(x)\varphi(x')\,G_{\mu\nu}(x,x')\nabla^\mu\nabla^{\prime\nu}\Delta(x,x')\Big\rbrack.\label{B22}
\end{align}
In an similar way
\begin{equation}
\langle S_2\rangle=i\int dv_x\Big\lbrack -\Big(\frac{e^2}{2\alpha}+\frac{\lambda}{3}\Big)\varphi^2(x)\,\Delta(x,x) + \frac{e^2}{2}\,\varphi^2(x)\,G^{\mu}{}_{\mu}(x,x)\Big\rbrack.\label{B23}
\end{equation}
The calculation of \eqref{B22} is the most involved so we will do it first.

To evaluate the pole parts of the products of Green functions we will use the local momentum space method of Bunch and Parker~\cite{BunchParker} and dimensional regularization~\cite{tHooftandVeltman}. An outline of the details is given in Appendix~\ref{appA}. Details of the extraction of the pole terms of the Green function expressions are given in Appendix~\ref{appB}.

If we use \eqref{B1.7a}, \eqref{B1.9} and \eqref{B1.16} in \eqref{B22} we obtain
\begin{align}
\textrm{PP}\left\lbrace\langle S_1^2\rangle\right\rbrace&=\frac{ie^2}{8\pi^2\epsilon}\,\int dv_x\Big\lbrace (\alpha-5)\,\nabla^\mu\varphi\nabla_\mu\varphi \nonumber\\
&+\Big(2-\alpha-\frac{1}{\alpha}\Big)\Big\lbrack m_s^2+\Big(\xi-\frac{1}{6}\Big)R\Big\rbrack\,\varphi^2\Big\rbrace.\label{B23a}
\end{align}
Using \eqref{B1.2} and \eqref{B1.3} in \eqref{B23} results in
\begin{align}
\textrm{PP}\left\lbrace\langle S_2\rangle\right\rbrace&=\frac{1}{8\pi^2\epsilon}\,\int dv_x\Big\lbrace \frac{e^2}{4}\,\Big(\frac{\alpha}{3}-1\Big)\,R\varphi^2 \nonumber\\
&-\Big(\frac{\lambda}{3}+\frac{e^2}{2\alpha}\Big)\Big\lbrack m_s^2+\Big(\xi-\frac{1}{6}\Big)R\Big\rbrack\,\varphi^2\Big\rbrace.\label{B24}
\end{align}
Using \eqref{B23a} and \eqref{B24} in \eqref{B20} we find that
\begin{align}
\textrm{PP}\left\lbrace \Gamma_2\right\rbrace&=\frac{1}{8\pi^2\epsilon}\,\int dv_x\Big\lbrace \frac{e^2}{2}(5-\alpha)\,\nabla^\mu\varphi\nabla_\mu\varphi+ {e^2}\,\Big(\frac{\alpha}{12}-\frac{1}{4}\Big)\,R\varphi^2 \nonumber\\
&-\Big\lbrack\frac{\lambda}{3}+e^2\Big(1-\frac{\alpha}{2}\Big)\Big\rbrack \Big\lbrack m_s^2+\Big(\xi-\frac{1}{6}\Big)R\Big\rbrack\,\varphi^2\Big\rbrace,\label{B25}
\end{align}
before the $\alpha\rightarrow0$ limit is taken. Note that the terms in $1/\alpha$ that occur separately in \eqref{B23a} and \eqref{B24} cancel to leave a result that is finite as $\alpha\rightarrow0$. This cancellation provides a useful check on the algebraic technicalities since the $\alpha\rightarrow0$ limit must exist. The final result for the pole terms in the effective action that are quadratic in $\varphi$ coming from the scalar and vector fields is (now letting $\alpha\rightarrow0$)
\begin{align}
\textrm{PP}\left\lbrace \Gamma_2\right\rbrace&=\frac{1}{8\pi^2\epsilon}\,\int dv_x\Big\lbrace \frac{5}{2}\,e^2\,\nabla^\mu\varphi\nabla_\mu\varphi-\frac{1}{4}\,e^2\,R\varphi^2 \nonumber\\
&-\Big(\frac{\lambda}{3}+e^2\Big)\Big\rbrack \Big\lbrack m_s^2+\Big(\xi-\frac{1}{6}\Big)R\Big\rbrack\,\varphi^2\Big\rbrace.\label{B26}
\end{align}

\subsection{$\textrm{PP}\left\lbrace \Gamma_4\right\rbrace$}\label{PPGamma4}

We will now use \eqref{B21} to evaluate the pole terms in the one-loop effective action that arise from the Bose fields and that are quartic in the background scalar field $\varphi$. We will take each of the four terms in \eqref{B21} in turn using \eqref{B15b} for $S_1$ and \eqref{B15c} for $S_2$. 

\subsubsection{$\textrm{PP}\left \lbrace \langle S_2^2\rangle\right\rbrace$}\label{sec2.2}

It is convenient to write \eqref{B15c} as the sum of the two terms,
\begin{subequations}\label{4.1}
\begin{align}
S_{21}&=-\int dv_x\,\varphi^2(x)\Big\lbrack \frac{\lambda}{4}\,\phi_1^2+\Big(\frac{\lambda}{12}+\frac{e^2}{2\alpha}\Big)\,\phi_2^2\Big\rbrack,\label{4.1a}\\
S_{22}&= \frac{e^2}{2}\int dv_x\,\varphi^2(x)\,A^\mu A_\mu.\label{4.1b}
\end{align}
\end{subequations}
It then follows that
\begin{equation}
\langle S_2^2\rangle=\langle S_{21}^2\rangle + \langle S_{22}^2\rangle.\label{4.2}
\end{equation}
The cross-term $\langle S_{21}S_{22}\rangle$ vanishes since it does not give rise to a connected Feynman diagram. Using \eqref{B17} it can be shown that
\begin{subequations}\label{4.3}
\begin{align}
\langle S_{21}^2\rangle&=-2\Big\lbrack \frac{\lambda^2}{16}+\Big(\frac{\lambda}{12}+\frac{e^2}{2\alpha}\Big)^2\Big\rbrack\int dv_x\int dv_{x'}\,\varphi^2(x)\,\varphi^2(x')\,\Delta^2(x,x'),\label{4.3a}\\
\langle S_{22}^2\rangle&= \frac{e^4}{2}\int dv_x\int dv_{x'}\,\varphi^2(x)\,\varphi^2(x')\,G^{\mu\nu}(x,x')G_{\mu\nu}(x,x').\label{4.3b}
\end{align}
\end{subequations}
Power counting shows that the pole parts of both $\Delta^2(x,x')$ and $G^{\mu\nu}(x,x')G_{\mu\nu}(x,x')$ come from the flat spacetime parts of the local momentum space expansions described in Appendix~\ref{appA}. From \eqref{B1.17} and \eqref{B1.18} we find
\begin{equation}\label{4.4}
\textrm{PP}\left \lbrace \langle  S_{2}^2\rangle\right\rbrace=-\,\frac{i}{8\pi^2\epsilon}\,\Big( \frac{5\lambda^2}{72}+\frac{\lambda\,e^2}{12\alpha}+\frac{e^4}{4\alpha^2}+\frac{3}{4}\,e^4+\frac{\alpha^2}{4}\,e^4\Big)\int dv_x\,\varphi^4(x),
\end{equation}
when the results from \eqref{4.3a} and \eqref{4.3b} are combined.

\subsubsection{$\textrm{PP}\left \lbrace \langle S_1^2S_2\rangle\right\rbrace$}\label{sec2.3}

We can write the result in \eqref{B15b} for $S_1$ as the sum of the two terms
\begin{subequations}\label{4.5}
\begin{align}
S_{11}&=e\,\Big(1+\frac{1}{\alpha}\Big)\int dv_x\,A^\mu(x)\,\phi_2(x)\,\nabla_{\!\!\mu}\varphi(x),\label{4.5a}\\
S_{12}&= - e\,\Big(1-\frac{1}{\alpha}\Big)\int dv_x\,\varphi(x)\,A^\mu(x)\,\nabla_{\!\!\mu}\phi_2(x).\label{4.5b}
\end{align}
\end{subequations}
Power counting shows that
\begin{equation}\label{4.6}
\textrm{PP}\left \lbrace \langle S_1^2S_2\rangle\right\rbrace = \textrm{PP}\left \lbrace \langle S_{12}^2S_2\rangle\right\rbrace
\end{equation}
If we use $S_2=S_{21}+S_{22}$ as in \eqref{4.1} we have
\begin{subequations}\label{4.7}
\begin{align}
\langle S_{12}^2S_{21}\rangle&=2ie^2\,\Big(1-\frac{1}{\alpha}\Big)^2\Big(\frac{\lambda}{12}+\frac{e^2}{2\alpha}\Big)\int dv_x\int dv_{x'}\int dv_{x''}\,\varphi(x)\varphi(x')\varphi^2(x'')\nonumber\\
&\qquad\times G^{\mu\nu}(x,x')\nabla_{\!\!\mu}\Delta(x,x'')\nabla_{\!\!\nu}^{\prime}\Delta(x',x''), 
\label{4.7a}\\
\langle S_{12}^2S_{22}\rangle&=-ie^4\,\Big(1-\frac{1}{\alpha}\Big)^2\int dv_x\int dv_{x'}\int dv_{x''}\,\varphi(x)\varphi(x')\varphi^2(x'')\nonumber\\
&\qquad\times G^{\mu\lambda}(x,x'')G^{\nu}{}_{\lambda}(x',x'')\nabla_{\!\!\mu}\nabla_{\!\!\nu}^{\prime}\Delta(x,x').\label{4.7b}
\end{align}
\end{subequations}
The pole parts of the Green's function expressions appearing in \eqref{4.7} are evaluated in \eqref{B1.19} and \eqref{B1.20} in Appendix~\ref{appB}. It can be shown that
\begin{equation}\label{4.8}
\textrm{PP}\left \lbrace -\frac{1}{2} \langle S_1^2S_2\rangle\right\rbrace = \frac{e^2\,(\alpha-1)^2}{16\pi^2\epsilon}\,\Big(\frac{\lambda}{6\alpha}+\frac{e^2}{\alpha^2}+e^2\Big)\int dv_x\,\varphi^4(x).
\end{equation}

\subsubsection{$\textrm{PP}\left \lbrace \langle S_1^4\rangle\right\rbrace$}\label{sec2.4}

If we use \eqref{4.5} it can be seen, based on power counting, that
\begin{align}
\textrm{PP}\left \lbrace \langle S_1^4\rangle\right\rbrace&=\textrm{PP}\left \lbrace \langle S_{12}^4\rangle\right\rbrace\nonumber\\
&=6e^4\,\Big(1-\frac{1}{\alpha}\Big)^4\int dv_x\int dv_{x'}\int dv_{x''}\int dv_{x'''}\,\varphi(x)\varphi(x')\varphi(x'')\varphi(x''')\nonumber\\
&\qquad\times G^{\mu\nu}(x,x')G^{\lambda\sigma}(x'',x''')\nabla_{\!\!\mu}\nabla_{\!\!\lambda}^{\prime\prime}\Delta(x,x'')\nabla_{\!\!\nu}^{\prime}\nabla_{\!\!\sigma}^{\prime\prime\prime}\Delta(x',x'''). \label{4.9}
\end{align}
The pole part of the product of Green's functions here is evaluated in \eqref{B1.21} from Appendix~\ref{appB} and gives
\begin{equation}\label{4.10}
\textrm{PP}\left \lbrace -\frac{i}{24} \langle S_1^4\rangle\right\rbrace = -\,\frac{{e^4}\,(\alpha-1)^4}{32\pi^2\epsilon\,{\alpha^2}}\int dv_x\,\varphi^4(x).
\end{equation}

If we now combine the three terms found in \eqref{4.4},\eqref{4.8}, and \eqref{4.10} the pole part of $\Gamma_4$ from \eqref{B21} turns out to be
\begin{equation}\label{4.11}
\textrm{PP}\left \lbrace \Gamma_4\right\rbrace = -\,\frac{1}{8\pi^2\epsilon} \Big\lbrack \frac{5\,}{72}\,\lambda^2+\frac{1}{6}\Big(1-\frac{\alpha}{2}\Big)\lambda e^2+\frac{5}{4}\,e^4 \Big\rbrack\int dv_x\,\varphi^4(x).
\end{equation}
All the potentially troublesome terms in $\alpha^{-1}$ and $\alpha^{-2}$ that arose at intermediate stages and which would have prevented taking the Landau-DeWitt limit $\alpha\rightarrow0$ have cancelled. The limit $\alpha\rightarrow0$ can now be taken in \eqref{4.11} to obtain the gauge independent result. It must be remembered that this is only part of the effective action and the ghost fields and fermions must also be included.

\subsection{Ghost contribution}\label{secghost}

From \eqref{B16} the ghost contribution to the one-loop effective action is
\begin{equation}
\Gamma^{(1)}_{\textrm{ghost}}=-i\ln\det(\Box+e^2\varphi^2).\label{G1.1}
\end{equation}
We will first use the perturbative approach utilized in the earlier sections to evaluate the pole part. The result will then be checked with the heat kernel method.

Start by writing \eqref{G1.1} as a functional integral over the Faddeev-Popov ghost fields $\bar{c}(x)$ and $c(x)$ which are treated as anticommuting:
\begin{equation}
\Gamma^{(1)}_{\textrm{ghost}}=-i\ln\int\lbrack dc\,d\bar{c}\rbrack\,e^{i\int dv_x\bar{c}(x)(\Box+e^2\varphi^2)c(x)}.\label{G1.2}
\end{equation}
We can treat the $e^2\varphi^2$ part as an interaction term,
\begin{equation}
S_{\textrm{ghost}}^{\textrm{int}} = e^2\,\int dv_x\,\varphi^2(x)\,\bar{c}(x)c(x).\label{G1.3}
\end{equation}
As in \eqref{B16} we find
\begin{equation}
\Gamma^{(1)}_{\textrm{ghost}}=-i\left\langle e^{i\,S_{\textrm{ghost}}^{\textrm{int}}}\right\rangle,\label{G1.4}
\end{equation}
with $\langle\cdots\rangle$ meaning to Wick reduce the expression with only connected terms kept. If we just concentrate on the terms that involve the background field $\varphi$ and that can contain poles we have
\begin{equation}
\textrm{PP}\left\lbrace\Gamma^{(1)}_{\textrm{ghost}} \right\rbrace=\textrm{PP}\left\lbrace\langle S_{\textrm{ghost}}^{\textrm{int}}\rangle + \frac{i}{2}\,\langle \left(S_{\textrm{ghost}}^{\textrm{int}}\right)^2\rangle  \right\rbrace.\label{G1.5}
\end{equation}
The Wick reduction is performed by treating $\bar{c},c$ as anticommuting with the basic relation
\begin{equation}
\langle c(x)\,\bar{c}(x')\rangle=-i\,\Delta_g(x,x'),\label{G1.6}
\end{equation}
where
\begin{equation}
-\Box\,\Delta_g(x,x')=\delta(x,x').\label{G1.7}
\end{equation}
The signs were chosen here so that $\Delta_g(x,x')$ coincides with the scalar field Green's function $\Delta(x,x')$ in \eqref{B18b} with $m_s^2=0$ and $\xi=0$. We immediately have the local momentum space expansion from \eqref{A1.14a} and \eqref{A1.14b} as
\begin{equation}
\Delta_g(x,x')=\int\frac{d^np}{(2\pi)^n}\,e^{ip\cdot y}\,\Big\lbrack \frac{1}{p^2}+\frac{2}{3}\,R^{\mu\nu}p_\mu p_\nu\,p^{-6}-\frac{1}{3}\,R\,p^{-4}+\cdots\Big\rbrack,\label{G1,8}
\end{equation}
where terms up to and including $p^{-4}$ have been shown.

Using \eqref{G1.3} and \eqref{G1.6} we have
\begin{subequations}\label{G1.9}
\begin{align}
\langle S_{\textrm{ghost}}^{\textrm{int}}\rangle&=ie^2\int dv_x\,\varphi^2(x)\,\Delta_g(x,x),\label{G1.9a}\\
\langle \left(S_{\textrm{ghost}}^{\textrm{int}}\right)^2\rangle&=e^4\int dv_x\int dv_{x'}\,\varphi^2(x)\,\varphi^2(x')\,\Delta_g(x,x')\,\Delta_g(x',x).\label{G1.9b}
\end{align}
\end{subequations}
Using the dimensionally regulated result of \eqref{B1.1a} and \eqref{B1.1b} it is easy to show that
\begin{subequations}\label{G1.10}
\begin{align}
\textrm{PP}\left\lbrace \Delta_g(x,x) \right\rbrace&=\frac{i}{48\pi^2\epsilon}\,R,\label{G1.10a}\\
\textrm{PP}\left\lbrace \Delta_g(x,x')\,\Delta_g(x',x) \right\rbrace&=-\frac{i}{8\pi^2\epsilon}\,\delta(x,x').\label{G1.10b}
\end{align}
\end{subequations}
The pole part of $\Gamma^{(1)}_{\textrm{ghost}}$ that depends on $\varphi$ is therefore given from \eqref{G1.5} by
\begin{equation}
\textrm{PP}\left\lbrace \Gamma^{(1)}_{\textrm{ghost}} \right\rbrace = \frac{1}{16\pi^2\epsilon}\int dv_x\left\lbrack -\frac{1}{3}\,e^2\,R\,\varphi^2+e^4\,\varphi^4\right\rbrack.\label{G1.11}
\end{equation}

As mentioned above we can use the heat kernel method to check this result. Use of known heat kernel coefficients \cite{DeWittdynamical,Gilkey75,Gilkey79} (see \cite{fulling1989aspects,avramidi2000heat,vassilevich2003heat,ParkerTomsbook,kirsten2010spectral} for reviews) allows us in addition to obtain the vacuum part of the pole part of the one-loop effective action coming from the ghost fields that is independent of the background scalar field. (The vacuum part could also be found by using the local momentum space method but this would entail working to higher order in the expansions than we have done here. See for example \cite{BunchParker,toms2014local}.) For any covariant derivative $D_\mu$ and any $Q(x)$ we have (using the notation of \cite[pages 193--194]{ParkerTomsbook})
\begin{equation}
\textrm{PP}\left\lbrace i\,\ln\det(D^2+Q)\right \rbrace=-\frac{1}{8\pi^2\epsilon}\int dv_x\,\textrm{tr}\,E_2(x),\label{G1.12}
\end{equation}
where
\begin{align}
E_2&=\left(\frac{1}{72}\,R^2-\frac{1}{180}\,R^{\mu\nu}R_{\mu\nu}+\frac{1}{180}\,R^{\mu\nu\lambda\sigma}R_{\mu\nu\lambda\sigma}\right)\,I\nonumber\\
&\quad +\frac{1}{12}\,W^{\mu\nu}W_{\mu\nu}+\frac{1}{2}\,Q^2-\frac{1}{6}\,R\,Q,\label{G1.13}
\end{align}
where $W_{\mu\nu}=\lbrack D_\mu,D_\nu\rbrack$. (A total derivative term that cannot contribute to \eqref{G1.12} has been omitted here.) For the ghosts $D_\mu=\nabla_\mu$ acting on scalars, so $W_{\mu\nu}=0$ for the ghost fields. There is only one field so the trace in \eqref{G1.12} is redundant. The expression for $Q$ is $Q=e^2\,\varphi^2(x)$ from \eqref{G1.1}. We therefore find
\begin{align}
\textrm{PP}\left\lbrace \Gamma^{(1)}_{\textrm{ghost}} \right\rbrace &= \frac{1}{16\pi^2\epsilon}\int dv_x\Big(\frac{1}{36}\,R^2-\frac{1}{90}\,R^{\mu\nu}R_{\mu\nu}+\frac{1}{90}\,R^{\mu\nu\lambda\sigma}R_{\mu\nu\lambda\sigma}\nonumber\\
&\quad -\frac{1}{3}\,e^2\,R\,\varphi^2+e^4\,\varphi^4\Big).\label{G1.14}
\end{align}
The terms that involve $\varphi$ are seen to be the same as those found earlier in \eqref{G1.11}. In addition to a vacuum contribution, the ghost fields will only contribute to the $\xi$ and $\lambda$ renormalization group functions.

\section{Contribution to the effective action from Fermi fields}\label{secFermi}

We now turn to the contributions from the fermion fields $\Psi$ and $\chi$ whose action was given in \eqref{B3}. This will be done in two ways, one using the part of the action that involves $\varphi$ treated as an interaction and proceeding as we did in Sec.~\ref{secBose}, and the other way using a functional approach.

\subsection{Perturbative approach}

From \eqref{B3} we can define a term in $S_1$ that is linear in $\varphi$ as $S_1=S_{11}+S_{12}$ where
\begin{align}
S_{11}&=-\,\frac{1}{\sqrt{2}}\int dv_x\,\varphi(x)\,\bar{\chi}(x)(w+iw_5\gamma_5)\Psi(x),\label{F1}\\
S_{12}&=-\,\frac{1}{\sqrt{2}}\int dv_x\,\varphi(x)\,\bar{\Psi}(x)(w^\ast+iw_5^\ast\gamma_5)\chi(x).\label{F2}
\end{align}
From \eqref{B20} and \eqref{B21} we have the parts of the one-loop effective action that are quadratic and quartic in $\varphi$ as
\begin{align}
\Gamma_2^{\textrm{fermion}}&=\frac{i}{2}\langle S_1^2\rangle,\label{F3}\\
\Gamma_4^{\textrm{fermion}}&=-\,\frac{i}{24}\langle S_1^4\rangle,\label{F4}
\end{align}
Note that there is no term in $S_2$ here, and that there are no terms odd in $\varphi$ as these would involve unequal (odd) numbers of $\Psi$ and of $\chi$ fields that integrate to zero in the functional integral. The essential difference between the Fermi and Bose cases is that here we must treat the fields $\Psi$ and $\chi$ as anticommuting in the functional integration.

We will define the Feynman Green's functions for the two spinor fields to be $\Psi(x,x')$ and $\chi(x,x')$ where
\begin{align}
\Big(i\gamma^\mu \nabla_{\!\!\mu}-m_{\psi}-im_{\psi5}\gamma_5\Big)\Psi(x,x')=-\delta(x,x'),\label{F5}\\
\Big(i\gamma^\mu\nabla_{\!\!\mu}-m_{\chi}-im_{\chi5}\gamma_5\Big)\chi(x,x')=-\delta(x,x').\label{F6}
\end{align}  
The basic results needed to evaluate \eqref{F3} and \eqref{F4} are
\begin{align}
\langle\Psi_\alpha(x)\bar{\Psi}_\beta(x')\rangle&=-i\Psi_{\alpha\beta}(x,x'),\label{F7}\\
\langle\chi_\alpha(x)\bar{\chi}_\beta(x')\rangle&=-i\chi_{\alpha\beta}(x,x'),\label{F8}
\end{align}
where $\alpha$ and $\beta$ denote spinor indices.

It is now straightforward to show that $\langle S_1^2\rangle=2\langle S_{11}S_{12}\rangle$ and then to show that
\begin{align}
\Gamma_2^{\textrm{fermion}}&=\frac{i}{2}\int dv_x\int dv_{x'}\,\varphi(x)\varphi(x')\nonumber\\
&\qquad\times\textrm{tr}\lbrack(w+iw_5\gamma_5)\Psi(x,x')(w^\ast+iw_5^\ast\gamma_5)\chi(x',x)\rbrack.\label{F9}
\end{align}
For \eqref{F4} it follows first that $\langle S_1^4\rangle=6\langle S_{11}^2S_{12}^2\rangle$, and then that
\begin{align}
\Gamma_4^{\textrm{fermion}}&=\frac{i}{8}\int dv_x\int dv_{x'}\int dv_{x''}\int dv_{x'''}\,\varphi(x)\varphi(x')\varphi(x'')\varphi(x''')\textrm{tr}\lbrack(w+iw_5\gamma_5)\Psi(x,x')\nonumber\\
&\times(w^\ast+iw_5^\ast\gamma_5)\chi(x',x'') (w+iw_5\gamma_5)\Psi(x'',x''')(w^\ast+iw_5^\ast\gamma_5)\chi(x''',x)\rbrack.\label{F10}
\end{align}

Before evaluating the pole parts of the two expressions in \eqref{F9} and \eqref{F10} we will show how they can be obtained using functional methods. This serves as a useful check on the results.

\subsection{Functional approach}

Write the fermion part of the action in \eqref{B3} in the matrix form
\begin{equation}
S_{\textrm{fermion}}=\int dv_x\int dv_{x'}\,\left(\bar{\Psi}(x),\bar{\chi}(x)\right) \left( \begin{array}{cc} A(x,x')&B(x,x')\\ C(x,x')&D(x,x')\end{array}\right)\left(\begin{array}{c} \Psi(x')\\ \chi(x')\end{array}\right),\label{F11}
\end{equation}
where
\begin{subequations}\label{F12}
\begin{align}
A(x,x')&=\Big(i\gamma^\mu \nabla_{\!\!\mu}-m_{\psi}-im_{\psi5}\gamma_5\Big)\delta(x,x'),\label{F12a}\\
B(x,x')&=-\,\frac{1}{\sqrt{2}}\,\varphi(x)\,(w+iw_5\gamma_5)\,\delta(x,x'),\label{F12b}\\
C(x,x')&=-\,\frac{1}{\sqrt{2}}\,\varphi(x)\,(w^\ast+iw_5^\ast\gamma_5)\,\delta(x,x'),\label{F12c}\\
D(x,x')&=\Big(i\gamma^\mu \nabla_{\!\!\mu}-m_{\chi}-im_{\chi5}\gamma_5\Big)\delta(x,x').\label{F12d}
\end{align}
\end{subequations}
Integration over the anticommuting fields $\Psi$ and $\chi$ gives the full contribution to the effective action coming from the fermions as
\begin{equation}
\Gamma_{\textrm{fermion}}=-i\,\ln\det\,{\mathbb F},\label{F13}
\end{equation}
where $\mathbb{F}$ is the matrix appearing in \eqref{F11}. The only $\varphi$ dependence is through $B$ and $C$ in \eqref{F12b} and \eqref{F12c}. We can write
\begin{equation}
\mathbb{F}=\left( \begin{array}{cc} A&0\\ 0&D\end{array}\right)\left\lbrack \left( \begin{array}{cc} I&0\\ 0&I\end{array}\right)+\left( \begin{array}{cc} 0&A^{-1}B\\ D^{-1}C&0\end{array}\right) \right\rbrack.\label{F15}
\end{equation}
Note that $(A^{-1}B)(x,x')=\int dv_{x''}\,A^{-1}(x,x'')B(x'',x')$ here. From \eqref{F5} it can be seen that
\begin{equation}
A^{-1}(x,x')=-\Psi(x,x'),\label{F16}
\end{equation}
and from \eqref{F6} that
\begin{equation}
D^{-1}(x,x')=-\chi(x,x'),\label{F17}
\end{equation}

Using \eqref{F15} in \eqref{F13} results in
\begin{equation}
\Gamma_{\textrm{fermion}}=-i\,\ln\det\left( \begin{array}{cc} A&0\\ 0&D\end{array}\right) -i\,{\textrm{Tr}}(I+X),\label{F18}
\end{equation}
where
\begin{equation}
X=\left( \begin{array}{cc} 0&A^{-1}B\\ D^{-1}C&0\end{array}\right).\label{F19}
\end{equation}
Here we use ${\textrm{Tr}}$ to denote the functional as well as the Dirac trace. So for example, ${\textrm{Tr}}\,X=\int dv_x\,{\textrm{tr}}\,X(x,x)$ where ${\textrm{tr}}$ is just the Dirac trace.

All the dependence on $\varphi$ occurs in $X$ in \eqref{F18}. The first term in \eqref{F18} gives the vacuum contribution that we will consider later. The term in ${\textrm{Tr}}(I+X)$ can be expanded in powers of $X$. Because $X$ takes the off-diagonal form given in \eqref{F19} all terms odd in $X$ will have a vanishing trace. This means that $\Gamma_{\textrm{fermion}}$ will be even in $\varphi$, a result that was also noted above using the perturbative approach. Keeping terms up to quartic order in $\varphi$ we have
\begin{equation}
\Gamma_{\textrm{fermion}}=-i\,\ln\det\left( \begin{array}{cc} A&0\\ 0&D\end{array}\right) +\frac{i}{2}\,{\textrm{Tr}}(X^2)+\frac{i}{4}\,{\textrm{Tr}}(X^4)+\cdots.\label{F20}
\end{equation}
From \eqref{F19} we have
\begin{subequations}\label{F21}
\begin{align}
X^2&= \left( \begin{array}{cc} A^{-1}BD^{-1}C&0\\ 0&D^{-1}CA^{-1}B\end{array}\right) ,\label{F21a}\\
X^4&= \left( \begin{array}{cc} (A^{-1}BD^{-1}C)^2&0\\ 0&(D^{-1}CA^{-1}B)^2\end{array}\right).\label{F21b}
\end{align}
\end{subequations}
It is now easy to see that
\begin{equation}
\frac{i}{2}\,{\textrm{Tr}}(X^2)=\Gamma_2^{\textrm{fermion}},\label{F22a}
\end{equation}
and that
\begin{equation}
\frac{i}{4}\,{\textrm{Tr}}(X^4)=\Gamma_4^{\textrm{fermion}},\label{F22b}
\end{equation}
where \eqref{F9} and \eqref{F10} are regained.

\subsection{Evaluation of $\textrm{PP}\left\lbrace\Gamma_2^{\textrm{fermion}}\right\rbrace$}

Suppose that we define
\begin{equation}\label{F23}
\zeta_2(x,x')=\textrm{PP}\left\lbrace \textrm{tr}\lbrack \Psi(x,x')(w^\ast+iw_5^\ast\gamma_5) \chi(x',x)(w+iw_5\gamma_5)\rbrack \right\rbrace.
\end{equation}
We need the local momentum space expansions for the Feynman Green's functions. The results follow from \cite{Tomsyukawa1}:
\begin{equation}
\Psi(x,x')=\int\frac{d^np}{(2\pi)^n}\,e^{ip\cdot y}\,\left\lbrack \Psi_0(p)+\Psi_1(p;x')+\cdots\right\rbrack,\label{F24}
\end{equation}
where
\begin{equation}\label{F25}
\Psi_0(p)=\frac{\pslash-m_\psi+im_{5\psi}\gamma_5}{(p^2-m_\psi^2-m_{5\psi}^2)}
\end{equation}
is the flat spacetime expression, and 
\begin{align}
\Psi_1(p;x^\prime)
&= \frac{1}{3}R_{\mu\alpha\lambda\beta}p^\mu p^\beta (\pslash-m_\psi+im_{5\psi}\gamma_5) \gamma^{\prime\lambda}\gamma^{\prime\alpha}(p^2-m_\psi^2-m_{5\psi}^2)^{-3}\nonumber\\
&\quad - \frac{1}{12}R(\pslash-m_\psi+im_{5\psi}\gamma_5)(p^2-m_\psi^2-m_{5\psi}^2)^{-2}\nonumber\\
&\quad+\frac{1}{2}R_{\mu\lambda}p^\mu(\pslash-m_\psi+im_{5\psi}\gamma_5)\gamma^{\prime\lambda}(\pslash-m_\psi+im_{5\psi}\gamma_5)(p^2-m_\psi^2-m_{5\psi}^2)^{-3},\label{F26}
\end{align}
contains all terms that are linear in the curvature. Power counting shows that the terms indicated are sufficient to calculate the pole part in \eqref{F23}. Similar expressions hold for the Feynman Green's function $\chi(x,x')$ that we will not indicate explicitly here.

It is easily seen that
\begin{equation}\label{F27}
\zeta_2(x,x')=\int\frac{d^np}{(2\pi)^n}\,e^{ip\cdot y}\,\left\lbrack \textrm{PP}\left\lbrace\zeta_{2\,\textrm{flat}}\right\rbrace+\textrm{PP}\left\lbrace\zeta_{2\,\textrm{curved}}\right\rbrace\right\rbrack,
\end{equation}
where 
\begin{align}
\zeta_{2\,\textrm{flat}}&=\int\frac{d^nq}{(2\pi)^n}\,\frac{\textrm{tr}\lbrack(\pslash+\qslash-m_\psi+im_{5\psi}\gamma_5)(w^\ast+iw_5^\ast\gamma_5)(\qslash-m_\chi+im_{5\chi}\gamma_5)(w+iw_5\gamma_5) \rbrack}{\lbrack (p+q)^2-m_\psi^2-m_{5\psi}^2\rbrack(q^2-m_\chi^2-m_{5\chi}^2)},\label{F28}\\
\zeta_{2\,\textrm{curved}}&=\int\frac{d^nq}{(2\pi)^n}\Big\lbrace \textrm{tr}\left\lbrack\Psi_0(p+q)(w^\ast+iw_5^\ast\gamma_5) \chi_1(q;x')(w+iw_5\gamma_5) \right\rbrack\nonumber\\
&\qquad\qquad\quad+\textrm{tr}\left\lbrack\Psi_1(p+q;x')(w^\ast+iw_5^\ast\gamma_5) \chi_0(q)(w+iw_5\gamma_5) \right\rbrack\Big\rbrace.\label{F29}
\end{align}
The pole parts are identified by expanding the integrands of \eqref{F28} and \eqref{F29} in powers of $q$ keeping those terms that behave like $q^{-4}$ for large $q$. After some calculation, using \eqref{B1.1}, it can be shown that
\begin{align}
\textrm{PP}\left\lbrace \zeta_{2\,\textrm{flat}}\right\rbrace =&-\,\frac{i}{2\pi^2\epsilon}\,(|w|^2+|w_5|^2)\left(-\frac{1}{2}p^2+m_\psi^2+m_{5\psi}^2+m_\chi^2+m_{5\chi}^2\right)\nonumber\\
&-\,\frac{i}{2\pi^2\epsilon}\,\lbrack(wm_\chi+w_5m_{5\chi})(w^\ast m_\psi+w_5^\ast m_{5\psi})\nonumber\\
&\qquad - (wm_{5\chi}-w_5m_{\chi})(w^\ast m_{5\psi}-w_5^\ast m_{\psi})\rbrack.\label{F30}
\end{align}
As with the Bose case we have
\begin{equation}
\int\frac{d^np}{(2\pi)^n}\,e^{ip\cdot y}\,p^2=-\,\Box_y\,\delta(y),\label{F31}
\end{equation}
and \eqref{B1.15} is used to return from Riemann normal to general coordinates.

The pole part of \eqref{F29} is simplified by noting that the $q^{-4}$ term comes from the $q^{-1}$ part of $\Psi_0$ or $\chi_0$, and the $q^{-3}$ part of $\chi_1$ or $\Psi_1$. This means that we may set $p=0$ in \eqref{F29} and ignore all the mass terms resulting in a reasonably simple calculation. The net result is
\begin{equation}
\textrm{PP}\left\lbrace \zeta_{2\,\textrm{curved}}\right\rbrace =\frac{i}{24\pi^2\epsilon}\,(|w|^2+|w_5|^2)\,R.\label{F32}
\end{equation}

Combining \eqref{F30} and \eqref{F32} in \eqref{F27} results in
\begin{align}
\textrm{PP}\left\lbrace \Gamma_{2}^{\textrm{fermion}}\right\rbrace &=\frac{1}{4\pi^2\epsilon}\int dv_x\Big\lbrace -\,\frac{1}{2}\,(|w|^2+|w_5|^2)\,\nabla^\mu\varphi\nabla_{\!\!\mu}\varphi+\frac{1}{12}\,\,(|w|^2+|w_5|^2)\,R\varphi^2\nonumber\\
&\qquad+\big\lbrack\,(|w|^2+|w_5|^2)(m_\psi^2+m_{5\psi}^2+m_\chi^2+m_{5\chi}^2)\nonumber\\
&\qquad +(wm_\chi+w_5m_{5\chi})(w^\ast m_\psi+w_5^\ast m_{5\psi})\nonumber\\
&\qquad + (wm_{5\chi}-w_5m_{\chi})(w_5^\ast m_{\psi}-w^\ast m_{5\psi})\big\rbrack\,\varphi^2\Big\rbrace.\label{F33}
\end{align}
This gives all terms in the one-loop effective action that are quadratic in $\varphi$ coming from the quantized fermions.

\subsection{Evaluation of $\textrm{PP}\left\lbrace\Gamma_4^{\textrm{fermion}}\right\rbrace$}

Power counting in \eqref{F10} shows that the pole part will come from just the flat spacetime terms in the local momentum space expansion. Furthermore, the fermion mass terms cannot contribute to the pole. After a bit of calculation, it can be shown that
\begin{equation}
\textrm{PP}\left\lbrace \Gamma_{4}^{\textrm{fermion}}\right\rbrace=\frac{1}{16\pi^2\epsilon}\big\lbrack (|w|^2+|w_5|^2)^2-(ww_5^\ast-w^\ast w_5)^2 \big\rbrack\int dv_x\,\varphi^4(x).\label{F34}
\end{equation}

We now have all terms arising from the fermion fields that can give rise to the scalar field renormalization, as well as the renormalization of the non-minimal coupling constant $\xi$, the scalar field mass, and the scalar field quartic coupling constant. The counterterms and renormalization group functions will be evaluated in the next section.

\section{Counterterms, renormalization group functions, and effective potential}\label{ctrg}

\subsection{Gravitational pole terms}

We can obtain the gravitational counterterms from the one-loop effective action by setting the background scalar field $\varphi=0$ and performing the functional integration. The gravitational part of the one-loop effective action is
\begin{align}
\Gamma_{\textrm{grav}}^{(1)}&=i\,\ln\det(\Box+m_s^2+\xi\,R)+\frac{i}{2}\,\ln\det\Big\lbrack\delta^{\mu}_{\nu}\Box+R^{\mu}_{\nu}-\Big(1-\frac{1}{\alpha}\Big)\,\nabla^\mu\nabla_\nu\Big\rbrack\nonumber\\
&\quad-i\,\ln\det(\Box)-i\,\ln\det(i\nablaslash-m_\psi-i\,m_{5\psi}\gamma_5)\nonumber\\
&\quad-i\,\ln\det(i\nablaslash-m_\chi-i\,m_{5\chi}\gamma_5).\label{RG1}
\end{align}
Here the $\alpha\rightarrow0$ limit is understood as we are using the Landau-DeWitt gauge. The first term arises from the two scalar degrees of freedom, the second one from the vector field, the third term from the ghosts, and the last two terms from the Dirac spinors. The basic heat kernel result (see \cite{ParkerTomsbook} for example) is that
\begin{equation}
{\textrm{PP}}\left\lbrace i\,\ln\det{\mathfrak O}\right\rbrace=-\,\frac{1}{8\pi^2\epsilon}\,\int dv_x\,{\textrm{tr}}E_2(x),\label{RG2}
\end{equation}
where ${\mathfrak O}$ is a second order differential operator like that in the first three terms of \eqref{RG1}, and $E_2(x)$ is a coefficient in the asymptotic expansion of the heat kernel for $\mathfrak{O}$. For reviews see \cite{DeWittdynamical,fulling1989aspects,avramidi2000heat,vassilevich2003heat,ParkerTomsbook,kirsten2010spectral}  for some of the literature. The most general derivation of $E_2$ was given by Gilkey~\cite{Gilkey75,Gilkey79} for the case of minimal operators (those whose leading second derivative terms involve only $\Box$). For non-minimal operators, like that for the vector field where $\nabla^\mu\nabla_\nu$ occurs, see \citep{BarvinskyVilkovisky} or \cite{MossToms} and references therein. The Dirac spinor contributions can be put into a second order from as in \cite{Tomsyukawa1} by defining a new covariant derivative. It can be shown that
\begin{equation}
\ln\det(i\nablaslash-m_0-im_5\gamma_5)=\frac{1}{2}\,\ln\det(D^2+Q),\label{RG3}
\end{equation}
where
\begin{subequations}
\begin{align}
D_\mu&=\nabla_\mu-m_5\gamma_5\gamma_\mu,\label{RG4}\\
Q&=\Big(m_0^2+3\,m_5^2+\frac{1}{4}\,R\Big)\,I+2i\,m_0\,m_5\,\gamma_5.\label{RG5}
\end{align}
\end{subequations}

For any operator of the form $\mathfrak{O}=D^2+Q$, the $E_2$ coefficient is given by \eqref{G1.13} 
%\begin{align}
%E_2&=\left(\frac{1}{72}\,R^2-\frac{1}{180}\,R^{\mu\nu}R_{\mu\nu}+\frac{1}{180}\,R^{\mu\nu\lambda\sigma}R_{\mu%\nu\lambda\sigma}\right)\,I\nonumber\\
%&\quad +\frac{1}{12}\,W^{\mu\nu}W_{\mu\nu}+\frac{1}{2}\,Q^2-\frac{1}{6}\,R\,Q,\label{RG6}
%\end{align}
where $W_{\mu\nu}=\lbrack D_\mu,D_\nu\rbrack$.
For the Dirac spinors, using \eqref{RG4} it follows that
\begin{equation}
W_{\mu\nu}=-\,\frac{1}{4}\,R_{\mu\nu\lambda\sigma}\,\gamma^\lambda\gamma^\sigma-m_5^2\,\lbrack\gamma_\mu,\gamma_\nu\rbrack.\label{RG8}
\end{equation}
This is sufficient information to evaluate the pole parts of all terms in \eqref{RG1} apart from that for the vector field. Due to the presence of the $\nabla^\mu\nabla_\nu$ term the operator is not of the form $D^2+Q$ where the result of \eqref{G1.13} can be applied. Operators where the covariant derivatives do not appear just in the form $D^2$ have been termed non-minimal by Barvinsky and Vilkovisky~\cite{BarvinskyVilkovisky} and they have developed a technique to deal with them. (See also \cite{gusynin1997computation,gusynin1999complete}.) The necessary $E_2$ coefficient for the real vector field has also been calculated using the local momentum expression in \cite{toms2014local} and more generally in \cite{MossToms}. It follows from these references that for the vector field operator that appears in \eqref{RG1} 
\begin{equation}
\textrm{PP}\Big\lbrace \ln\det\Big\lbrack\delta^{\mu}_{\nu}\Box+R^{\mu}_{\nu}-\Big(1-\frac{1}{\alpha}\Big)\,\nabla^\mu\nabla_\nu\Big\rbrack\Big\rbrace = \textrm{PP}\Big\lbrace \ln\det\Big(\delta^{\mu}_{\nu}\Box+R^{\mu}_{\nu}\Big)\Big\rbrace,\label{RG8b}
\end{equation}
provided that terms that are total derivatives are discounted. (This is not true if $R^\mu_\nu$ is replaced with something else, or if the total derivatives are included in the $E_2$ coefficient; however, we only require the integrated $E_2$ coefficient here.) 

The pole terms in $\Gamma_{\textrm{grav}}^{(1)}$ can now be shown to be
\begin{align}
\textrm{PP}\Big\lbrace \Gamma_{\textrm{grav}}^{(1)}\Big\rbrace &= -\,\frac{1}{16\pi^2\epsilon}\int dv_x\Big\lbrace m_s^4-2\,(m_\psi^2+m_{\psi5}^2)^2-2\,(m_\chi^2+m_{\chi5}^2)^2\nonumber\\
&\qquad+\Big\lbrack 2\Big(\xi-\frac{1}{6}\Big)\,m_s^2-\frac{1}{3}\,(m_{\psi}^2+ m_{\psi5}^2 + m_{\chi}^2 + m_{\chi}^2)\Big\rbrack\,R\nonumber\\
&\qquad-\frac{1}{45}\,R^{\mu\nu\lambda\sigma}R_{\mu\nu\lambda\sigma} + \frac{47}{90}\,R^{\mu\nu}R_{\mu\nu} + \Big(\frac{2}{3}\,\xi^2-\frac{5}{36}\Big)R^2\Big\rbrace.\label{RG9}
\end{align}

\subsection{Counterterms}

The bare classical action follows from \eqref{2.2} as (keeping only the background scalar field $\Phi=\varphi/\sqrt{2}$ and gravitational field non-zero) as 
\begin{align}
S&=\int dv_x \Big( \frac{1}{2}\,\nabla^\mu\varphi_{\textrm{B}}\nabla_\mu\varphi_{\textrm{B}} -\frac{1}{2}\,m_{\textrm{s\,B}}^2\,\varphi_{\textrm{B}}^2 - \frac{1}{2}\, \xi_{\textrm{B}}\,R \,\varphi_{\textrm{B}}^2 -\frac{\lambda_{\textrm{B}}}{4!}\,\varphi_{\textrm{B}}^4 \nonumber\\
&\qquad+ \Lambda_{\textrm{B}}+\kappa_{\textrm{B}}\,R+\alpha_{1\,\textrm{B}}\,R^{\mu\nu\lambda\sigma}R_{\mu\nu\lambda\sigma} + \alpha_{2\,\textrm{B}}\,R^{\mu\nu}R_{\mu\nu} + \alpha_{3\,\textrm{B}}\,R^2 \Big),\label{RG10}
\end{align}
with the subscript `$B$' denoting a bare quantity. We will define the renormalization counterterms, following `t~Hooft~\cite{tHooft1973}, by
\begin{subequations}\label{RG11}
\begin{align}
\varphi_\bare&=\mu^{\epsilon/2}(1+\delta Z_\varphi)\varphi,\label{RG11a}\\
m_{s\,\bare}^2&=m_s^2+\delta m_s^2,\label{RG11b}\\
\xi_\bare&=\xi+\delta\xi,\label{RG11c}\\
\lambda_\bare&=\mu^{-\epsilon}(\lambda+\delta\lambda),\label{RG11d}\\
\Lambda_{\bare}&=\mu^{\epsilon}(\Lambda+\delta\Lambda),\label{RG11e}\\
\kappa_{\bare}&=\mu^{\epsilon}(\kappa+\delta\kappa),\label{RG11f}\\
\alpha_{i\,\bare}&=\mu^{\epsilon}(\alpha_i+\delta\alpha_i).\ i=1,2,3.\label{RG11g}
\end{align}
\end{subequations}
The `t~Hooft unit of mass $\mu$ gives the renormalized quantities the dimensions for all $n$ that they have in the physical spacetime dimension $n=4$. 

The counterterm part of the action that will be used to absorb the one-loop pole terms coming from the full effective action will be
\begin{align}
S_{\textrm{ct}}&=\int dv_x \Big\lbrack \delta Z_\varphi\,\nabla^\mu\varphi\nabla_\mu\varphi -\Big(\frac{1}{2}\,\delta m_{\textrm{s}}^2+m_{\textrm{s}}^2\,\delta Z_\varphi\Big)\varphi^2 - \Big(\frac{1}{2}\,\delta \xi+\xi\,\delta Z_\varphi\Big)\,R \,\varphi^2\label{RG12}\\
&\qquad -\Big(\frac{\delta\lambda}{4!}+\frac{\lambda}{6}\,\delta Z_\varphi\Big)\varphi^4 + \delta\Lambda+\delta\kappa\,R+\delta\alpha_{1}\,R^{\mu\nu\lambda\sigma}R_{\mu\nu\lambda\sigma} + \delta\alpha_{2}\,R^{\mu\nu}R_{\mu\nu} + \delta\alpha_{3}\,R^2 \Big\rbrack.\nonumber
\end{align}

The counterterms in \eqref{RG12} are fixed by requiring that $S_{\textrm{ct}}+\textrm{PP}\lbrace \Gamma^{(1)}\rbrace$ remain finite as $\epsilon\rightarrow0$. If all the pole terms calculated previously in \eqref{B26},\eqref{4.11},\eqref{G1.14},\eqref{F33},\eqref{F34}, and \eqref{RG9} are combined it can be seen that 
\begin{subequations}\label{RG13}
\begin{align}
\delta Z_\varphi&=\frac{1}{16\pi^2\epsilon}\,(2\,|w|^2+2\,|w_5|^2-5\,e^2),\label{RG13a}\\
\delta m_{s}^2&=\frac{1}{4\pi^2\epsilon}\Big\lbrack \Big(\frac{3}{2}\,e^2-\frac{\lambda}{3}-|w|^2-|w_5|^2\Big)\,m_s^2 \nonumber\\
&\qquad+2\,(|w|^2+|w_5|^2)(m_\psi^2+m_{5\psi}^2+m_\chi^2+m_{5\chi}^2)\nonumber\\
&\qquad+2\,(|w|^2-|w_5|^2)(m_\psi m_\chi-m_{5\psi} m_{5\chi})\nonumber\\
&\qquad +2\,(ww_5^\ast+w^\ast w_5)(m_\psi m_{5\chi}+m_{5\psi} m_{\chi})  \Big\rbrack\,,\label{RG13b}\\
\delta \xi&=\frac{1}{24\pi^2\epsilon}\,(9\,e^2-2\,\lambda-6\,|w|^2-6\,|w_5|^2)\Big(\xi-\frac{1}{6}\Big),\label{RG13c}\\
\delta\lambda&=\frac{1}{24\pi^2\epsilon}\,\lbrack -5\,\lambda^2+18\,\lambda\,e^2-54\,e^4-12\,\lambda\,(|w|^2+|w_5|^2)\nonumber\\
&\qquad +36\,(|w|^2+|w_5|^2)^2-36\,(ww_5^\ast-w^\ast w_5)^2\rbrack,\label{RG13d}\\
\delta\Lambda&=-\,\frac{1}{16\pi^2\epsilon}\,\lbrack2\,(m_\psi^2+m_{5\psi}^2)^2 + 2\,(m_\chi^2+m_{5\chi}^2)^2 - m_s^4\rbrack,\label{RG13e}\\
\delta\kappa&=-\,\frac{1}{48\pi^2\epsilon}\,\lbrack m_\psi^2+m_{5\psi}^2+m_\chi^2+m_{5\chi}^2-(6\xi-1)m_s^2\rbrack,\label{RG13f}\\
\delta\alpha_{1}&=-\,\frac{1}{480\pi^2\epsilon},\label{RG13g}\\
\delta\alpha_{2}&=-\,\frac{1}{30\pi^2\epsilon},\label{RG13h}\\
\delta\alpha_{3}&=\frac{1}{48\pi^2\epsilon}\,\Big(\xi^2-\frac{1}{4}\Big).\label{RG13i}
\end{align}
\end{subequations}

\subsection{Renormalization group functions}

It is now possible to apply `t~Hooft's method~\cite{tHooft1973} to calculate the renormalization group functions from the counterterms. We follow the notation and conventions of \cite{ParkerTomsbook} with $q_i$ representing any of the terms entering the theory, including the background field $\varphi$. The change in $q_i$ under a change in the renormalization mass scale $\mu$ is given by
\begin{equation}
\mu\,\frac{d}{d\mu}\,q_i=\beta_{q_i}.\label{RG14}
\end{equation}
The renormalization group functions are found from the one-loop counterterms given in \eqref{RG13} to be
\begin{subequations}\label{RG15}
\begin{align}
\beta_\varphi&=\frac{1}{16\pi^2}\,(5\,e^2-2\,|w|^2-2\,|w_5|^2)\,\varphi,\label{RG15a}\\
\beta_{m_{s}^2}&=\frac{1}{4\pi^2}\Big\lbrack \Big(\frac{\lambda}{3}+|w|^2+|w_5|^2-\frac{3}{2}\,e^2)\,m_s^2 -2\,(|w|^2+|w_5|^2)(m_\psi^2+m_{5\psi}^2+m_\chi^2+m_{5\chi}^2)\nonumber\\
&\qquad-2\,(|w|^2-|w_5|^2)(m_\psi m_\chi-m_{5\psi} m_{5\chi}) -2\,(ww_5^\ast+w^\ast w_5)(m_\psi m_{5\chi}+m_{5\psi} m_{\chi})  \Big\rbrack\,,\label{RG15b}\\
\beta_{\xi}&=\frac{1}{24\pi^2}\,(2\,\lambda-9\,e^2+6\,|w|^2+6\,|w_5|^2)\Big(\xi-\frac{1}{6}\Big),\label{RG15c}\\
\beta_{\lambda}&=\frac{1}{24\pi^2}\,\lbrack 5\,\lambda^2-18\,\lambda\,e^2+54\,e^4+12\,\lambda\,(|w|^2+|w_5|^2)\nonumber\\
&\qquad -36\,(|w|^2+|w_5|^2)^2+36\,(ww_5^\ast-w^\ast w_5)^2\rbrack,\label{RG15d}\\
\beta_{\Lambda}&=\frac{1}{16\pi^2}\,\lbrack2\,(m_\psi^2+m_{5\psi}^2)^2 + 2\,(m_\chi^2+m_{5\chi}^2)^2 - m_s^4\rbrack,\label{RG15e}\\
\beta_{\kappa}&=\frac{1}{48\pi^2}\,\lbrack m_\psi^2+m_{5\psi}^2+m_\chi^2+m_{5\chi}^2-(6\xi-1)m_s^2\rbrack,\label{RG15f}\\
\beta_{\alpha_{1}}&=-\,\frac{1}{480\pi^2},\label{RG15g}\\
\beta_{\alpha_{2}}&=-\,\frac{1}{30\pi^2},\label{RG15h}\\
\beta_{\alpha_{3}}&=\frac{1}{48\pi^2}\,\Big(\xi^2-\frac{1}{4}\Big).\label{RG15i}
\end{align}
\end{subequations}

\subsection{Effective action}

In the case where there are no mass scales present in the classical theory (apart from the fields) the method of Coleman and Weinberg~\cite{ColemanandWeinberg} can be used to evaluate the terms in the effective action in terms of the renormalization group functions. We will only be concerned with what is obtained at one-loop order, rather than the exact results given in \citep{ColemanandWeinberg}, often referred to as renormalization group improved. The method described in \cite{Tomsyukawa1} can be used to show that
\begin{equation}
\Gamma=\int dv_x\Big\lbrack \frac{1}{2}Z(\varphi)\partial^\mu\varphi\partial_\mu\varphi-V_0(\varphi)-RV_1(\varphi)+\alpha_{1}(\varphi)R^{\mu\nu\lambda\sigma}R_{\mu\nu\lambda\sigma} + \alpha_{2}(\varphi)R^{\mu\nu}R_{\mu\nu} + \alpha_{3}(\varphi)R^2 \Big\rbrack,\label{RG16}
\end{equation}
where $Z(\varphi),V_0(\varphi),V_1(\varphi),\alpha_i(\varphi)$ are given to one-loop order by
\begin{subequations}\label{RG17}
\begin{align}
Z(\varphi)&=1+A\,\ln(\varphi^2/\mu^2),\label{RG17a}\\
V_0(\varphi)&=\frac{\lambda}{4!}\,\varphi^4+B\,\varphi^4\Big\lbrack \ln(\varphi^2/\mu^2)-\frac{25}{6}\Big\rbrack,\label{RG17b}\\
V_1(\varphi)&=\frac{1}{2}\,\xi\,\varphi^2+C\,\varphi^2\lbrack \ln(\varphi^2/\mu^2)-3\rbrack,\label{RG17c}\\
\alpha_i(\varphi)&=\alpha_i+D_i\, \ln(\varphi^2/\mu^2),\label{RG17d}
\end{align}
\end{subequations}
where
\begin{subequations}\label{RG18}
\begin{align}
A&={\beta}_{\varphi}/\varphi\nonumber\\
&=\frac{1}{16\pi^2}(5\,e^2-2\,|w|^2-2\,|w_5|^2),\label{RG18a}\\
B&=\frac{1}{48}\,\beta_\lambda+\frac{\lambda}{12\,\varphi}\,{\beta}_{\varphi}\nonumber\\
&= \frac{1}{192\pi^2}\Big\lbrack\frac{5}{6}\,\lambda^2+2\,\lambda e^2+9\,e^4-6\,(|w|^2+|w_5|^2)^2 +6\,(ww_5^\ast-w^\ast w_5)^2\Big\rbrack,\label{RG18b}\\
C&=\frac{1}{4}\,\beta_\xi+\frac{1}{2\,\varphi}\,\xi\,\tilde{\beta}_{\varphi}\nonumber\\
&= \frac{1}{192\pi^2}\Big\lbrack(4\,\lambda+12\,e^2 )\big(\xi-1/6\big)+(5\,e^2-2\,|w|^2-2\,|w_5|^2)\Big\rbrack,\label{RG18c}
\end{align}
\end{subequations}
and $D_i=\frac{1}{2}\beta_{\alpha_i}$ so that
\begin{subequations}
\begin{align}
\alpha_1(\varphi)&=\alpha_1-\frac{1}{960\pi^2}\,\ln(\varphi^2/\mu^2),\label{RG19a}\\
\alpha_2(\varphi)&=\alpha_2-\frac{1}{60\pi^2}\,\ln(\varphi^2/\mu^2),\label{RG19b}\\
\alpha_3(\varphi)&=\alpha_1+\frac{1}{96\pi^2}\,(\xi^2-1/4)\,\ln(\varphi^2/\mu^2).\label{RG19c}
\end{align}
\end{subequations}
This gives a complete evaluation of those terms in the one-loop effective action that can be found from renormalization group considerations. The results have been established in a way that respects gauge invariance, independence of the choice of gauge condition, and also in a way that is independent of the choice made for the scalar field parametrization.

\section{Conclusions and discussion}

We have considered the one-loop counterterms for a charged scalar field interacting with a gauge field and Dirac spinors through a Yukawa interaction. These counterterms were used to calculate the renormalization group functions and the curved spacetime effective potential up to and including order $R^2$ along with the gradient terms in the scalar field in the massless case. The background scalar field was not assumed to be constant so that the field renormalization could be calculated. All calculations were done in a way that respects gauge and field parametrization invariance, and crucially independence from the choice of gauge condition. The local momentum space method was used along with some heat kernel results. We did not present a full analysis of all one-loop counterterms and renormalization group functions as was done in the simpler case \cite{Tomsyukawa1}, but the methods used there could be applied here without any essential difficulties. 

We included an unconventional pseudoscalar mass term of the generic form $m_5\bar{\psi}\gamma_5\psi$ for each of the two Dirac spinors in \eqref{2.2c}. As noted in \cite{Tomsyukawa1} this term can be transformed away in flat spacetime by a chiral rotation of the Dirac fields. However, in curved spacetime there is an anomaly and the effective action is not invariant under the necessary transformation. For the theory in the present paper a similar analysis to that presented in \cite{Tomsyukawa1} shows that the change in the effective action under the necessary transformation is
\begin{equation}
\Delta\Gamma=-\frac{1}{768\pi^2}\,\left\lbrack \tan^{-1}\left(\frac{m_{5\psi}}{m_\psi}\right)+  \tan^{-1}\left(\frac{m_{5\chi}}{m_\chi}\right)\right\rbrack\int dv_x\,\epsilon^{\lambda\sigma\rho\tau}R_{\mu\nu\lambda\sigma}R^{\mu\nu}{}_{\rho\tau}.\label{CD1}
\end{equation}
An outline of the calculation is given in Appendix~\ref{secanomaly}. The details of the calculation are essentially the same as those that appear in the axial, or chiral, anomaly and are most easily seen using the path integral method of Fujikawa~\cite{Fujikawa1,Fujikawa2,Fujikawa3}. The only difference is in the overall coefficient here that involves the two possible mass terms. The integral is seen to involve the Pontryagin density just as in the axial anomaly. This conclusion holds also if a background vector field is included as seen in Appendix~\ref{secanomaly}. There is still some current interest in such expressions. (See for example, \cite{Bonoraetal}.)

This result in \eqref{CD1} is exact. The transformations necessary to remove the pseudoscalar mass terms, given in \eqref{C1a} and \eqref{C1b}, also change the coefficients in the Yukawa interaction. There is also the option of transforming away either the scalar or else the pseudoscalar Yukawa interactions instead of the pseudoscalar mass term. Again, an anomaly like that in \eqref{CD1} will result with $w$ and $w_5$ appearing in place of the masses. Because the anomaly term is finite and is independent of the quantized fields it cannot affect the perturbative evaluation of the counterterms. 

It is possible to generalize the analysis that we have presented here to the non-Abelian case. It is also possible to work in a more general choice of gauge and see exactly how the gauge condition parameters disappear from the effective action if the Vilkovisky-DeWitt formalism is used as was done in the pioneering calculation of Fradkin and Tseytlin~\cite{FradkinTseytlin}. The details are somewhat more involved than those presented here and will be given elsewhere.

\appendix\section{Local momentum space expansions}\label{appA}

Consider the Green function $G^{i}{}_{j}(x,x')$ where $i$ and $j$ refer to any type of indices ({\em eg.\/} vector or tensor). For the case of spacetime indices it is advantageous to refer them to a local orthonormal frame by using the vierbein formalism as noted in \cite{toms2014local}. Suppose that the Green's function obeys
\begin{equation}
\left\lbrack \left( A^{\mu\nu}\right)^{i}{}_{j}\partial_\mu\partial_\nu +\left(B^\mu\right)^{i}{}_{j}\partial_\mu+C^{i}{}_{j}\right\rbrack\,G^{j}{}_{k}(x,x')=\delta^{i}_{k}\delta(x,x').\label{A1.1}
\end{equation}
Here $A^{\mu\nu},B^\mu$ and $C$ are some functions of $x$ that are specific to the Green's function being considered. They will be specified for scalars in \eqref{A1.9} and for vectors in \eqref{A1.10} below.

The basic idea behind the local momentum space method~\cite{BunchParker} is to introduce Riemann normal coordinates at the point in spacetime whose local coordinates are $x^{\prime\mu}$ and to expand about that point using
\begin{equation}
x^\mu=x^{\prime\mu}+y^\mu.\label{A1.2}
\end{equation}
Expressions for $A^{\mu\nu},B^\mu$ and $C$ are developed as a power series in $y^\mu$. We will take (suppressing the indices $i$ and $j$ here)
\begin{subequations}\label{A1.3}
\begin{align}
A^{\mu\nu}(x)&=A^{\mu\nu}(x')+A^{\mu\nu}{}_{\alpha\beta}\,y^\alpha y^\beta+\cdots,\label{A1.3a}\\
B^{\mu}(x)&=B^{\mu}{}_{\alpha}\,y^\alpha +\cdots,\label{A1.3b}\\
C(x)&=C(x')+C_{\alpha}\,y^\alpha+\cdots,\label{A1.3c}
\end{align}
\end{subequations}
The absence of a linear term in \eqref{A1.3a} and a zeroth order term in \eqref{A1.3b} will be seen to hold in our case but the method does not rely on either of these assumptions. For the Green's function we take
\begin{equation}
G^{i}{}_{j}(x,x')=\int \frac{d^np}{(2\pi)^n}\,e^{ip\cdot y}\,G^{i}{}_{j}(p;x'),\label{A1.4}
\end{equation}
where $G^{i}{}_{j}(p;x')$ can depend on the origin of the Riemann normal coordinates. We can expand $G^{i}{}_{j}(p;x')$ as an asymptotic series in $p$ whose coefficients depend on the terms in the expansions given in \eqref{A1.3}. If we write
\begin{equation}
G^{i}{}_{j}(p;x')=G_{0}{}^{i}{}_{j}(p;x')+G_{2}{}^{i}{}_{j}(p;x')+\cdots,\label{A1.5}
\end{equation}
where the subscript $0,2,\ldots$ counts the dimension (in units of mass or inverse length) of the coefficient of $p$, it can be shown that~\cite{toms2014local}
\begin{equation}
-\left(A^{\mu\nu}(x')\right)^{i}{}_{j}\,p_\mu p_\nu\,G_{0}{}^{j}{}_{k}=\delta^{i}_{k},\label{A1.6}
\end{equation}
and that
\begin{equation}
G_{2}{}^{i}{}_{j}(p;x')=G_{21}{}^{i}{}_{j}(p;x')+G_{22}{}^{i}{}_{j}(p;x')+G_{23}{}^{i}{}_{j}(p;x'),\label{A1.7}
\end{equation}
where
\begin{subequations}\label{A1.8}
\begin{align}
G_{21}{}^{i}{}_{j}(p;x')&=-G_{0}{}^{i}{}_{k}(p;x')\,\left( A^{\mu\nu}{}_{\alpha\beta}\right)^{k}{}_{l}\, \frac{\partial^2}{\partial p_\alpha\partial p_\beta}\left\lbrack p_\mu p_\nu\, G_{0}{}^{l}{}_{j}(p;x')\right\rbrack, \label{A1.8a}\\
G_{22}{}^{i}{}_{j}(p;x')&= G_{0}{}^{i}{}_{k}(p;x')\,\left( B^{\mu}{}_{\alpha}\right)^{k}{}_{l}\, \frac{\partial}{\partial p_\alpha}\left\lbrack p_\mu\, G_{0}{}^{l}{}_{j}(p;x')\right\rbrack, \label{A1.8b}\\
G_{23}{}^{i}{}_{j}(p;x')&= -G_{0}{}^{i}{}_{k}(p;x')\,\left( C(x')\right)^{k}{}_{l}\, G_{0}{}^{l}{}_{j}(p;x'). \label{A1.8c}
\end{align}
\end{subequations}
Because it follows from \eqref{A1.6} that $G_0\sim p^{-2}$ for large $p$ from \eqref{A1.8} we can conclude that $G_2\sim p^{-4}$ for large $p$. Higher order terms in the expansion \eqref{A1.5} will fall off even faster than $p^{-4}$. This means that we will not need any of the higher order terms in our calculation.

The scalar field Green's function obeys \eqref{B18b}. By comparison with \eqref{A1.1} we can identify (leaving off the spacetime coordinates)
\begin{subequations}\label{A1.9}
\begin{align}
A^{\mu\nu}&=-g^{\mu\nu},\label{A1.9a}\\
B^\mu&=g^{\lambda\sigma}\Gamma^{\mu}_{\lambda\sigma},\label{A1.9b}\\
C&=-m_s^2-\xi R.\label{A1.9c}
\end{align}
\end{subequations}

For the vector field, from \eqref{B18a} we have
\begin{subequations}\label{A1.10}
\begin{align}
\left(A^{\mu\nu}\right)^{a}{}_{b}&=\delta^{a}_{b}g^{\mu\nu}-\Big(1-\frac{1}{\alpha}\Big) (e^{a\mu}e_{b}{}^{\nu} +e^{a\nu}e_{b}{}^{\mu}),\label{A1.10a}\\
\left(B^\mu\right)^{a}{}_{b}&=2g^{\mu\nu}\,\omega_{\nu}{}^{a}{}_{b}-\delta^{a}_{b} g^{\lambda\sigma} \Gamma^{\mu}_{\lambda\sigma}+\Big(1-\frac{1}{\alpha}\Big)e^{a\lambda}e_{b}{}^{\sigma}\Gamma^{\mu}_{\lambda\sigma}\nonumber\\
&-\Big(1-\frac{1}{\alpha}\Big)(e^{a\mu}e_{c}{}^{\nu} + e^{a\nu}e_{c}{}^{\mu})\,\omega_{\nu}{}^{c}{}_{b},\label{A1.10b}\\
\left(C\right)^{a}{}_{b}&=g^{\mu\nu}\partial_\mu\,\omega_{\nu}{}^{a}{}_{b}- g^{\lambda\sigma} \Gamma^{\mu}_{\lambda\sigma}\,\omega_{\mu}{}^{a}{}_{b} + g^{\mu\nu}\,\omega_{\mu}{}^{a}{}_{c}\,\omega_{\nu}{}^{c}{}_{b}\nonumber\\
&+R^{a}{}_{b}- \Big(1-\frac{1}{\alpha}\Big)e^{a\mu}e_{c}{}^{\nu}\partial_\mu\omega_{\nu}{}^{c}{}_{b}\nonumber\\
&+\Big(1-\frac{1}{\alpha}\Big)e^{a\mu}e_{c}{}^{\nu}\Gamma^{\lambda}_{\mu\nu}\,\omega_{\lambda}{}^{c}{}_{b} - \Big(1-\frac{1}{\alpha}\Big)e^{a\mu}e_{c}{}^{\nu}\,\omega_{\mu}{}^{c}{}_{d}\,\omega_{\nu}{}^{d}{}_{b}.\label{A1.10c}
\end{align}
\end{subequations}
Here we use $a,b,c,d$ to denote orthonormal frame indices with the vierbein $e^{a}{}_{\mu}$ defined as usual by
\begin{equation}
g_{\mu\nu}=e^{a}{}_{\mu}e^{b}{}_{\nu}\,\eta_{ab}.\label{A1.11}
\end{equation}
$\omega_{\mu}{}^{a}{}_{b}$ is the spin connection for the vector field which is given by~\cite[page 223]{ParkerTomsbook}
\begin{equation}
\omega_{\mu}{}^{a}{}_{b}=-e_{b}{}^{\nu}(\partial_\mu e^{a}{}_{\nu}-\Gamma^{\lambda}_{\mu\nu}e^{a}{}_{\lambda}).\label{A1.12}
\end{equation}
Spacetime indices are raised and lowered with the spacetime metric $g_{\mu\nu}$ and orthonormal frame indices are raised and lowered with $\eta_{ab}$. The expansions of the metric, vierbein and connections in Riemann normal coordinates that we require are
\begin{subequations}\label{A1.13}
\begin{align}
g_{\mu\nu}(x)&=\eta_{\mu\nu}+\frac{1}{3}\,R_{\mu\alpha\nu\beta}\,y^\alpha y^\beta+\cdots,\label{A1.13a}\\
g^{\mu\nu}(x)&=\eta^{\mu\nu}-\frac{1}{3}\,R^{\mu}{}_{\alpha}{}^{\nu}{}_{\beta}\,y^\alpha y^\beta+\cdots,\label{A1.13b}\\
\Gamma^{\lambda}_{\mu\nu}(x)&=\frac{1}{3}\,\left( R^{\lambda}{}_{\mu\nu\alpha}+R^{\lambda}{}_{\nu\mu\alpha}\right)\,y^\alpha+\cdots,\label{A1.13c}\\
e^{a}{}_{\mu}(x)&=e^{a}{}_{\lambda}(x')\left(\delta^{\lambda}_{\mu}+\frac{1}{6}\,R^{\lambda}{}_{\alpha\mu\beta}\,y^\alpha y^\beta+\cdots\right),\label{A1.13d}\\
e_{a}{}^{\mu}(x)&=e_{a}{}^{\lambda}(x')\left(\delta_{\lambda}^{\mu}-\frac{1}{6}\,R^{\mu}{}_{\alpha\lambda\beta}\,y^\alpha y^\beta+\cdots\right),\label{A1.13e}\\
\omega_{\mu}{}^{a}{}_{b}(x)&=\frac{1}{2}\,R^{a}{}_{b\mu\alpha}\,y^\alpha +\cdots.\label{A1.13f}
\end{align}
\end{subequations}
All curvature terms on the right-hand side of \eqref{A1.13} are evaluated at the origin of Riemann normal coordinates $x'$. Note that $R^{a}{}_{b\mu\alpha}=e^{a}{}_{\lambda}(x')e_{b}{}^{\sigma}(x')R^{\lambda}{}_{\sigma\mu\alpha}(x') $ in \eqref{A1.13f}.

By substituting \eqref{A1.13} into \eqref{A1.9} and \eqref{A1.10} we can find the expressions required to evaluate the Green's function expansion terms in \eqref{A1.6}. After some calculation it can be shown that
\begin{subequations}\label{A1.14}
\begin{align}
\Delta_0(p)&=\frac{1}{p^2},\label{A1.14a}\\
\Delta_2(p;x')&=\left(\xi-\frac{1}{3}\right)\,R\,p^{-4}+m_s^2\,p^{-4}+\frac{2}{3}\,R^{\mu\nu}\,p_\mu p_\nu\, p^{-6},\label{A1.14b}\\
G_{0}{}^{a}{}_{b}(p)&=-\delta^{a}_{b}\,p^{-2}+(1-\alpha)\,p^a p_b\, p^{-4},\label{A1.14c}\\
G_{2}{}^{a}{}_{b}(p;x')&=\frac{1}{3}\,\delta^{a}_{b}\,R\,p^{-4}+\frac{2}{3}\,(\alpha-1)\,R\,p^ap_b\,p^{-6} + \frac{1}{6}\,(\alpha-7)\,R^{a}{}_{b}\,p^{-4}\nonumber\\
&\quad-\frac{2}{3}\,\delta^{a}_{b}\,R^{\mu\nu}\,p_\mu p_\nu\,p^{-6}+2(1-\alpha)\,R^{\mu\nu}\,p_\mu p_\nu p^a p_b\,p^{-8}\nonumber\\
&\quad+\frac{2}{3}\,(1-\alpha)\,R^{a\mu}{}_{b}{}^{\nu}\,p_\mu p_\nu\,p^{-6}.\label{A1.14d}
\end{align}
\end{subequations}
These give the terms in the Green's function expansion that we will need. They agree with those in \cite{toms2014local} where the higher order terms in the expansion for the vector field Green's function can be found.

\section{Pole parts of products of Green functions}\label{appB}

In this Appendix we will describe how the results found for the Green's functions in Appendix~\ref{appA} can be used to evaluate the pole terms in the products of Green functions needed in our evaluation of the pole part of the one-loop effective action. Because we are using dimensional regularization it suffices to evaluate momentum space integrals whose integrands behave like $p^{-4}$ for large $p$. This avoids the necessity of combining denominators using Feynman or Schwinger parameters. The basic integrals needed are 
\begin{subequations}\label{B1.1}
\begin{align}
{\textrm{PP}}\left\lbrace \int\frac{d^np}{(2\pi)^n}\,\frac{1}{p^4} \right\rbrace&=-\frac{i}{8\pi^2\epsilon},\label{B1.1a}\\
{\textrm{PP}}\left\lbrace \int\frac{d^np}{(2\pi)^n}\,\frac{p_\mu p_\nu}{p^6} \right\rbrace&=-\frac{i}{32\pi^2\epsilon}\,\eta_{\mu\nu},\label{B1.1b}\\
{\textrm{PP}}\left\lbrace \int\frac{d^np}{(2\pi)^n}\,\frac{p_\mu p_\nu p_\lambda p_\sigma}{p^8} \right\rbrace&=-\frac{i}{192\pi^2\epsilon}\,(\eta_{\mu\nu}\eta_{\lambda\sigma} + \eta_{\mu\lambda}\eta_{\nu\sigma} + \eta_{\mu\sigma}\eta_{\lambda\nu}).\label{B1.1c}
\end{align}
\end{subequations}

From \eqref{A1.14b} it can be seen that
\begin{equation}
{\textrm{PP}}\left\lbrace \Delta(x,x) \right\rbrace=-\,\frac{i}{8\pi^2\epsilon}\Big\lbrack m_s^2+\Big(\xi-\frac{1}{6}\Big)R\Big\rbrack.\label{B1.2}
\end{equation}
This result can also be obtained from the known coefficients in the heat kernel expansion as described originally in \cite{TomsPRDscalar} and provides a check on the local momentum space expansion.

From \eqref{A1.14d} it can be shown that
\begin{equation}
{\textrm{PP}}\left\lbrace G^{\mu}{}_{\mu}(x,x) \right\rbrace=-\frac{i}{8\pi^2\epsilon}\Big( \frac{\alpha}{6}-\frac{1}{2}\Big)R.\label{B1.3}
\end{equation}
This result can also be obtained from the known heat kernel coefficient for nonminimal operators as found in \cite{toms2014local} or \cite{MossToms}.

Turning next to $\Delta(x,x')G^{a}{}_{b}(x,x')$ we have upon using the local momentum space expansions \eqref{A1.4} for each Green function
\begin{equation}
\Delta(x,x')G^{a}{}_{b}(x,x')=\int\frac{d^np}{(2\pi)^n}\,e^{ip\cdot y}\,F^{a}{}_{b}(p;x'),\label{B1.4}
\end{equation}
where
\begin{equation}
F^{a}{}_{b}(p;x')=\int\frac{d^nq}{(2\pi)^n}\,\Delta(p-q;x')G^{a}{}_{b}(q;x').\label{B1.5}
\end{equation}
Only terms in the integrand of \eqref{B1.5} that behave like $q^{-4}$ for large $q$ will result in a pole, so it is clear that the flat spacetime expressions \eqref{A1.14a} and \eqref{A1.14c} can be used here. The result is
\begin{equation}
{\textrm{PP}}\left\lbrace F^{a}{}_{b}(p;x') \right\rbrace=\frac{i}{32\pi^2\epsilon}\,(\alpha+3)\,\delta^{a}_{b}, \label{B1.6}
\end{equation}
giving
\begin{equation}
{\textrm{PP}}\left\lbrace \Delta(x,x')G^{a}{}_{b}(x,x') \right\rbrace=\frac{i}{32\pi^2\epsilon}\,(\alpha+3)\,\delta^{a}_{b}\,\delta(y). \label{B1.7}
\end{equation}
The presence of the Dirac $\delta$ on the right-hand side of \eqref{B1.7} allows us to deduce that
\begin{equation}
{\textrm{PP}}\left\lbrace \Delta(x,x')G^{\mu\nu}(x,x') \right\rbrace=\frac{i}{32\pi^2\epsilon}\,(\alpha+3)\,g^{\mu\nu}(x)\,\delta(x,x'), \label{B1.7a}
\end{equation}
upon the return to general coordinates.

We also need $\textrm{PP}\left\lbrace G^{\mu\nu}(x,x')\nabla_{\!\!\mu}\Delta(x,x') \right\rbrace$. We can use the local momentum space expansions for the Green functions to write
\begin{equation}
\nabla_{\!\!\mu}\Delta(x,x')G^{a}{}_{b}(x,x')=\int\frac{d^np}{(2\pi)^n}\,e^{ip\cdot y}\int\frac{d^nq}{(2\pi)^n}\,i(p_\mu-q_\mu)\Delta(p-q;x')G^{a}{}_{b}(q;x').\label{B1.8}
\end{equation}
Power counting again shows that the pole term coming from the integrand in \eqref{B1.8} that behaves like $q^{-4}$ can be found using the flat spacetime terms \eqref{A1.14a} and \eqref{A1.14c}. After some calculation, and returning to general coordinates, it follows that
\begin{equation}
\textrm{PP}\left\lbrace G^{\mu\nu}(x,x')\nabla_{\!\!\mu}\Delta(x,x') \right\rbrace=-\frac{i}{32\pi^2\epsilon}\,(\alpha-3)\,\nabla^{\nu}\delta(x,x').\label{B1.9}
\end{equation}

The last term needed for the evaluation of $\langle S_1^2\rangle$ in \eqref{B22} is $\textrm{PP}\left\lbrace G^{\mu\nu}(x,x')\nabla_{\!\!\mu}\nabla^{\prime}_{\!\!\nu}\Delta(x,x') \right\rbrace$. The calculation of this expression is a bit more involved than the previous ones. We can write
\begin{align}
F(x,x')&=G^{\mu\nu}(x,x')\nabla_{\!\!\mu}\nabla^{\prime}_{\!\!\nu}\Delta(x,x')\nonumber\\
&=F_{a}{}^{b}(x,x')G^{a}{}_{b}(x,x'),\label{B1.10}
\end{align}
with
\begin{equation}
F_{a}{}^{b}(x,x')=e_{a}{}^{\mu}(x)e^{b\nu}(x')\nabla_{\!\!\mu}\nabla^{\prime}_{\!\!\nu}\Delta(x,x').\label{B1.11}
\end{equation}
By using the expansion \eqref{A1.13e} and that for the scalar field Green function \eqref{A1.4} along with \eqref{A1.14a} and \eqref{A1.14b} it follows that 
\begin{align}
F_{a}{}^{b}(x,x')&=\int\frac{d^np}{(2\pi)^n}\,e^{ip\cdot y}\Big\lbrack p^ap_b\,\Delta_1(p;x') +p^a p_b\,p^{-2}\nonumber\\
&\qquad-\frac{1}{6}\,R^{a}{}_{b}\,p^{-2} - \frac{1}{3}\,R^{a\mu\nu}{}_{b}\,p_\mu p_\nu\,p^{-4}+\cdots\Big\rbrack,\label{B1.12}
\end{align}
where the higher order terms not shown fall off faster than $p^{-2}$. Because the vector field Green function behaves at least like $p^{-2}$ these higher order terms cannot contribute to the pole part of $G^{\mu\nu}(x,x')\nabla_{\!\!\mu}\nabla^{\prime}_{\!\!\nu}\Delta(x,x')$. We then find that
\begin{equation}
F(x,x')=e_{a}{}^{\mu}(x')e^{b\nu}(x')\int\frac{d^np}{(2\pi)^n}\,e^{ip\cdot y}F^{a}{}_{b\mu\nu}(p;x'),\label{B1.13a}
\end{equation}
where
\begin{align}
F^{a}{}_{b\mu\nu}(p;x')&=\int\frac{d^np}{(2\pi)^n}\,G^{a}{}_{b}(q;x')\,\Big\lbrack \frac{(p_\mu-q_\mu)(p_\nu-q_\nu)}{(p-q)^2} - \frac{1}{6}\,R_{\mu\nu}\,(p-q)^{-2}\nonumber\\
&\quad-\frac{1}{3}\,R_{\mu}{}^{\lambda\sigma}{}_{\nu}\,(p_\lambda-q_\lambda)(p_\sigma-q_\sigma)(p-q)^{-4}\nonumber\\
&\quad + (p_\mu-q_\mu)(p_\nu-q_\nu)\Delta_1(p-q;x')+\cdots\Big\rbrack.\label{B1.13}
\end{align}
The integrand of \eqref{B1.13} can now be expanded in powers of $q$ keeping terms that behave like $q^{-4}$. Both \eqref{A1.14c} and \eqref{A1.14d} must be used here. After some calculation it follows that
\begin{equation}
\textrm{PP}\left\lbrace F(x,x')\right\rbrace = -\frac{i}{8\pi^2\epsilon}\,\Big\lbrack \frac{3}{4}\,(1-\alpha)\,\Box_y-\alpha\, m_s^2+\Big(\frac{5}{12}\,\alpha-\frac{1}{4}-\alpha\,\xi\Big)\,R\Big\rbrack\,\delta(y).\label{B1.14}
\end{equation}
To return from Riemann normal to general coordinates we must use~\cite{Tomsyukawa1} 
\begin{equation}
\Big(\Box_x+\frac{1}{3}\,R\Big)\,\delta(x,x')=\Box_y\,\delta(y).\label{B1.15}
\end{equation}
This leads to the result that
\begin{equation}
\textrm{PP}\left\lbrace G^{\mu\nu}(x,x')\nabla_{\!\!\mu}\nabla^{\prime}_{\!\!\nu}\Delta(x,x') \right\rbrace=  -\frac{i}{8\pi^2\epsilon}\,\Big\lbrace \frac{3}{4}\,(1-\alpha)\,\Box_x - \alpha\,\Big\lbrack m_s^2+\Big(\xi-\frac{1}{6}\Big)\,R\Big\rbrack\Big\rbrace\,\delta(x,x').\label{B1.16}
\end{equation}

In Sec.~\ref{secBose} we need $\textrm{PP}\left\lbrace\Delta^2(x,x')\right\rbrace$. This is easily evaluated using the flat spacetime part of the local momentum space expansion given in \eqref{A1.14a}. The curvature term in \eqref{A1.14b} cannot contribute to the pole part. It is easily shown that
\begin{align}
\textrm{PP}\left\lbrace\Delta^2(x,x')\right\rbrace&=\textrm{PP}\left\lbrace\int\frac{d^np}{(2\pi)^n}\,e^{ip\cdot y}\int\frac{d^nq}{(2\pi)^n}\,\Delta_0(p-q)\Delta_0(q)\right\rbrace\nonumber\\
&=-\frac{i}{8\pi^2\epsilon}\,\delta(x,x').\label{B1.17}
\end{align}

In a similar way by using \eqref{A1.14c} it can be shown that
\begin{align}
\textrm{PP}\left\lbrace G^{\mu\nu}(x,x')G_{\mu\nu}(x,x')\right\rbrace&=\textrm{PP}\left\lbrace\int\frac{d^np}{(2\pi)^n}\,e^{ip\cdot y}\int\frac{d^nq}{(2\pi)^n}\,G_0^{\mu\nu}(p-q)G_{0\mu\nu}(q)\right\rbrace\nonumber\\
&=-\frac{i}{8\pi^2\epsilon}\,(3+\alpha^2)\,\delta(x,x').\label{B1.18}
\end{align}

In \eqref{4.7a} we require $\textrm{PP}\left\lbrace G^{\mu\nu}(x,x')\nabla_{\!\!\mu}\Delta(x,x'')\nabla_{\!\!\nu}^{\prime}\Delta(x',x'')\right\rbrace$. Power counting shows that the curvature corrections to the Green's functions cannot contribute to the pole coming from the product of Green's functions. The calculation is therefore identical to the flat spacetime result and it is easy to show that 
\begin{equation}
\textrm{PP}\left\lbrace G^{\mu\nu}(x,x')\nabla_{\!\!\mu}\Delta(x,x'')\nabla_{\!\!\nu}^{\prime}\Delta(x',x'')\right\rbrace= \frac{i\,\alpha}{8\pi^2\epsilon}\,\delta(x,x'')\,\delta(x',x'').\label{B1.19}
\end{equation}

In a similar way the expression needed in \eqref{4.7b} can be shown to be
\begin{equation}
\textrm{PP}\left\lbrace G^{\mu\lambda}(x,x'')G^{\nu}{}_{\lambda}(x',x'')\nabla_{\!\!\mu}\nabla_{\!\!\nu}^{\prime}\Delta(x,x')\right\rbrace= - \frac{i\,\alpha^2}{8\pi^2\epsilon}\,\delta(x,x')\,\delta(x',x'').\label{B1.20}
\end{equation}

To evaluate the product of Green's functions in \eqref{4.9} it can again be shown that the pole terms come only from the flat spacetime expansions in the local momentum space expressions. It can be shown that
\begin{align}
\textrm{PP}\left\lbrace G^{\mu\nu}(x,x')G^{\lambda\sigma}(x'',x''')\nabla_{\!\!\mu}\nabla_{\!\!\lambda}^{\prime\prime}\Delta(x,x'')\nabla_{\!\!\nu}^{\prime}\nabla_{\!\!\sigma}^{\prime\prime\prime}\Delta(x',x''')\right\rbrace&\nonumber\\
&\hspace{-24pt}= - \frac{i\,\alpha^2}{8\pi^2\epsilon}\,\delta(x,x'')\,\delta(x'',x''')\,\delta(x',x''').\label{B1.21}
\end{align}

\section{The anomaly}\label{secanomaly}

We outline the main steps in the derivation that leads up to \eqref{CD1} in this appendix. The path integral approach of Fujikawa~\cite{Fujikawa1,Fujikawa2,Fujikawa3} is used here and we follow our earlier paper~\cite{Tomsyukawa1}. We also make use of \cite[Sec.~5.9]{ParkerTomsbook} for some of the intermediate details.

The spinor fields $\chi$ and $\Psi$ in \eqref{2.2c} can be transformed as
\begin{align}
\chi(x)&=e^{-i\,\vartheta\,\gamma_5}\,\chi^\prime(x),\label{C1a}\\
\Psi(x)&=e^{-i\,\omega\,\gamma_5}\,\Psi^\prime(x),\label{C1b}
\end{align}
where the angles $\vartheta$ and $\omega$ are chosen to eliminate the pseudoscalar mass terms. Specifically we choose
\begin{subequations}\label{C2}
\begin{align}
\sin(2\vartheta)&=\frac{m_{\chi5}}{(m_\chi^2+m_{\chi5}^2)^{1/2}},\label{C2a}\\
\cos(2\vartheta)&=\frac{m_\chi}{(m_\chi^2+m_{\chi5}^2)^{1/2}},\label{C2b}\\
\sin(2\omega)&=\frac{m_{\psi5}}{(m_\psi^2+m_{\psi5}^2)^{1/2}},\label{C2c}\\
\cos(2\omega)&=\frac{m_\psi}{(m_\psi^2+m_{\psi5}^2)^{1/2}},\label{C2d}
\end{align}
\end{subequations}
The Yukawa terms in \eqref{2.2c} will also transform but we do not require the explicit form of this here. The classical theories based on the original and transformed fields will be identical. However, there will be an anomaly in the quantum theory~\cite{Tomsyukawa1} due to the parity violating pseudoscalar mass terms.

To calculate this anomaly it is expedient to adopt Fujikawa's~~\cite{Fujikawa1,Fujikawa2,Fujikawa3} method and analyse the change in the measure of the functional integral for the fermion part of the theory. If we let $\chi_N$ be a complete orthonormal set of solutions to the Dirac equation from \eqref{2.2c},
\begin{equation}
(i\gamma^\mu\nabla_\mu-m_{\chi}-im_{\chi5}\gamma_5)\chi_N(x)=\lambda_N\chi_N(x),\label{C3}
\end{equation}
and similarly let $\psi_N$ be a complete orthonormal set of solutions to
\begin{equation}
(i\gamma^\mu D_\mu-m_{\psi}-im_{\psi5}\gamma_5)\psi_N(x)=\widetilde{\lambda_N}\psi_N(x),\label{C4}
\end{equation}
then the effective action for the transformed fields $\chi^\prime$ and $\Psi^\prime$, that we will call $\Gamma^\prime$, is related to the original effective action $\Gamma$ for the original fields $\chi$ and $\Psi$ by
\begin{equation}
\Gamma^\prime=\Gamma+2i\ln\det\,C_{NN^\prime}+2i\ln\det\,\widetilde{C}_{NN^\prime}.\label{C5}
\end{equation}
The expressions $C_{NN^\prime}$ and $\widetilde{C}_{NN^\prime}$ come from the Jacobians in the functional measure under \eqref{C1a} and \eqref{C1b}. The explicit expressions are
\begin{align}
C_{NN^\prime}&=\cos\vartheta\,\delta_{NN^\prime}+i\mu\sin\vartheta\int dv_x\bar{\chi}_N(x)\gamma_5\chi_{N^\prime}(x).\label{C6a}\\
\widetilde{C}_{NN^\prime}&=\cos\omega\,\delta_{NN^\prime}+i\mu\sin\omega\int dv_x\bar{\psi}_N(x)\gamma_5\psi_{N^\prime}(x).\label{C6b}
\end{align}
We will allow there to be a background vector field present in the covariant derivative in \eqref{C4} for generality although this is not central to the calculation. 

By making use of the orthonormality and completeness of the modes $\chi_N$ and $\psi_N$ it can be shown as described in \cite{Tomsyukawa1} that
\begin{align}
\ln\det\,C_{NN^\prime}&=-\frac{\vartheta}{16\pi^2}\int dv_x{\textrm{tr}}\lbrack\gamma_5\,E_2(x)\rbrack,\label{C7a}\\
\ln\det\,\widetilde{C}_{NN^\prime}&=-\frac{\omega}{16\pi^2}\int dv_x{\textrm{tr}}\lbrack\gamma_5\,\widetilde{E}_2(x)\rbrack,\label{C7b}
\end{align}
where $E_2$ and $\widetilde{E}_2$ are the heat kernel coefficients for the Dirac operators in \eqref{C3} and \eqref{C4}.
Making use of \eqref{G1.13} shows that
\begin{align}
\ln\det\,C_{NN^\prime}&=\frac{i\,\vartheta}{768\pi^2}\,\int dv_x\,\epsilon^{\lambda\sigma\rho\tau}R_{\mu\nu\lambda\sigma}R^{\mu\nu}{}_{\rho\tau},\label{C8a}\\
\ln\det\,\widetilde{C}_{NN^\prime}&=\frac{i\,\omega}{768\pi^2}\,\int dv_x\,\epsilon^{\lambda\sigma\rho\tau}R_{\mu\nu\lambda\sigma}R^{\mu\nu}{}_{\rho\tau} - \frac{i\,e^2\,\omega}{32\pi^2}\,\int dv_x\,\epsilon^{\mu\nu\lambda\sigma}F_{\mu\nu}F_{\lambda\sigma}.\label{C8b}
\end{align}
(More details of the derivation can be found in \cite[Sec.~5.9]{ParkerTomsbook}.)

Substitution of \eqref{C8a} and \eqref{C8b} back into \eqref{C5} shows that
\begin{equation}
\Gamma^\prime=\Gamma-\frac{(\vartheta+\omega)}{384\pi^2}\,\int dv_x\,\epsilon^{\lambda\sigma\rho\tau}R_{\mu\nu\lambda\sigma}R^{\mu\nu}{}_{\rho\tau}+\frac{e^2\,\omega}{16\pi^2}\,\int dv_x\,\epsilon^{\mu\nu\lambda\sigma}F_{\mu\nu}F_{\lambda\sigma}.\label{C9}
\end{equation}
$\vartheta$ and $\omega$ can be eliminated in terms of the masses using \eqref{C3}. This leads directly to \eqref{CD1} if the background vector field is dropped.
%%%%%

%\bibliography{yukawa}

\begin{thebibliography}{99}
\bibitem{Tomsyukawa1}
D.~J. {Toms}, \emph{{Effective action for the Yukawa model in curved
  spacetime}}, \href{https://doi.org/10.1007/JHEP05(2018)139}{\emph{Journal of
  High Energy Physics} {\bfseries 5} (2018) 139}
  [\href{https://arxiv.org/abs/1804.08350}{{\ttfamily 1804.08350}}].

\bibitem{Vilkovisky1}
G.~A. Vilkovisky, \emph{Quantum theory of gravity},  ch.~{The Gospel according
  to DeWitt}, pp.~169--209.
\newblock Adam Hilger, 1984.

\bibitem{DeWitt6}
B.~S. DeWitt, \emph{{Relativity, Groups and Topology II}},  ch.~The spacetime
  approach to quantum field theory, pp.~381--738.
\newblock North Holland, 1984.

\bibitem{shapiro1989asymptotic}
I.~L. Shapiro, \emph{Asymptotic behaviour of effective {Y}ukawa coupling
  constants in quantum ${R}^2$ gravity with matter},
  \href{https://doi.org/10.1088/0264-9381/6/8/019}{\emph{Classical and Quantum
  Gravity} {\bfseries 6} (1989) 1197}.

\bibitem{odintsov1992general}
S.~D. Odintsov and I.~L. Shapiro, \emph{General relativity as the low-energy
  limit in higher derivative quantum gravity},
  \href{https://doi.org/10.1088/0264-9381/9/4/006}{\emph{Classical and Quantum
  Gravity} {\bfseries 9} (1992) 873}.

\bibitem{elizalde1994renormalization}
E.~Elizalde and S.~D. Odintsov, \emph{Renormalization-group improved effective
  potential for interacting theories with several mass scales in curved
  spacetime}, \href{https://doi.org/10.1007/BF01957780}{\emph{Zeitschrift
  f{\"u}r Physik C Particles and Fields} {\bfseries 64} (1994) 699}.

\bibitem{elizalde1995higgs}
E.~Elizalde and S.~D. Odintsov, \emph{Higgs-{Y}ukawa model in curved
  spacetime}, \href{https://doi.org/10.1103/PhysRevD.51.5950}{\emph{Physical
  Review D} {\bfseries 51} (1995) 5950}.

\bibitem{elizalde1995improved}
E.~Elizalde, S.~D. Odintsov and A.~Romeo, \emph{Improved effective potential in
  curved spacetime and quantum matter--higher derivative gravity theory},
  \href{https://doi.org/10.1103/PhysRevD.51.1680}{\emph{Physical Review D}
  {\bfseries 51} (1995) 1680}.

\bibitem{elizalde1995renormalization}
E.~Elizalde and S.~D. Odintsov, \emph{A renormalization group improved nonlocal
  gravitational effective {L}agrangian},
  \href{https://doi.org/10.1142/S0217732395001952}{\emph{Modern Physics Letters
  A} {\bfseries 10} (1995) 1821}.

\bibitem{ProkopecWoodard}
T.~{Prokopec} and R.~P. {Woodard}, \emph{{Production of massless fermions
  during inflation}},
  \href{https://doi.org/10.1088/1126-6708/2003/10/059}{\emph{Journal of High
  Energy Physics} {\bfseries 10} (2003) 059}
  [\href{https://arxiv.org/abs/astro-ph/0309593}{{\ttfamily
  astro-ph/0309593}}].

\bibitem{Garbrechtfermion}
B.~{Garbrecht} and T.~{Prokopec}, \emph{{Fermion mass generation in de Sitter
  space}}, \href{https://doi.org/10.1103/PhysRevD.73.064036}{\emph{Physical
  Review D} {\bfseries 73} (2006) 064036}
  [\href{https://arxiv.org/abs/gr-qc/0602011}{{\ttfamily gr-qc/0602011}}].

\bibitem{Garbrecht}
B.~{Garbrecht}, \emph{{Ultraviolet regularization in de Sitter space}},
  \href{https://doi.org/10.1103/PhysRevD.74.043507}{\emph{Physical Review D}
  {\bfseries 74} (2006) 043507}
  [\href{https://arxiv.org/abs/hep-th/0604166}{{\ttfamily hep-th/0604166}}].

\bibitem{Miao}
S.~P. {Miao} and R.~P. {Woodard}, \emph{{Leading log solution for inflationary
  Yukawa theory}},
  \href{https://doi.org/10.1103/PhysRevD.74.044019}{\emph{Physical Review D}
  {\bfseries 74} (2006) 044019}
  [\href{https://arxiv.org/abs/gr-qc/0602110}{{\ttfamily gr-qc/0602110}}].

\bibitem{Shapiro2011PLB}
F.~{Sobreira}, B.~J. {Ribeiro} and I.~L. {Shapiro}, \emph{{Effective potential
  in curved space and cut-off regularizations}},
  \href{https://doi.org/10.1016/j.physletb.2011.10.016}{\emph{Physics Letters
  B} {\bfseries 705} (2011) 273}
  [\href{https://arxiv.org/abs/1107.2262}{{\ttfamily 1107.2262}}].

\bibitem{herranen2014spacetime}
M.~{Herranen}, T.~{Markkanen}, S.~{Nurmi} and A.~{Rajantie}, \emph{{Spacetime
  Curvature and the Higgs Stability During Inflation}},
  \href{https://doi.org/10.1103/PhysRevLett.113.211102}{\emph{Physical Review
  Letters} {\bfseries 113} (2014) 211102}
  [\href{https://arxiv.org/abs/1407.3141}{{\ttfamily 1407.3141}}].

\bibitem{Cz}
O.~{Czerwinska}, Z.~{Lalak} and L.~{Nakonieczny}, \emph{{Stability of the
  effective potential of the gauge-less top-Higgs model in curved spacetime}},
  \href{https://doi.org/10.1007/JHEP11(2015)207}{\emph{Journal of High Energy
  Physics} {\bfseries 11} (2015) 207}
  [\href{https://arxiv.org/abs/1508.03297}{{\ttfamily 1508.03297}}].

\bibitem{markkanen20181}
T.~Markkanen, S.~Nurmi, A.~Rajantie and S.~Stopyra, \emph{{The 1-loop effective
  potential for the Standard Model in curved spacetime}}, {\emph{{ArXiv High
  Energy Physics e-prints}} (2018) }
  [\href{https://arxiv.org/abs/1804.02020}{{\ttfamily 1804.02020}}].

\bibitem{ColemanandWeinberg}
S.~Coleman and E.~Weinberg, \emph{Radiative corrections as the origin of
  spontaneous symmetry breaking},
  \href{https://doi.org/10.1103/PhysRevD.7.1888}{\emph{Physical Review D}
  {\bfseries 7} (1973) 1888}.

\bibitem{BuchbinderandOdintsovRG}
I.~L. Buchbinder and S.~D. Odintsov, \emph{Effective potential and phase
  transitions induced by curvature in gauge theories in curved spacetime},
  \href{https://doi.org/10.1088/0264-9381/2/5/014}{\emph{Class. and Quantum
  Grav.} {\bfseries 2} (1985) 721}.

\bibitem{ParkerTomsbook}
L.~E. Parker and D.~J. Toms, \emph{Quantum Field Theory in Curved Spacetime}.
  Cambridge University Press, 2009.

\bibitem{TomsRG}
D.~J. Toms, \emph{The effective action and the renormalization group equation
  in curved spacetime},
  \href{https://doi.org/10.1016/0370-2693(83)90011-4}{\emph{Physics Letters B}
  {\bfseries 126} (1983) 37}.

\bibitem{Zanusso}
O.~{Zanusso}, L.~{Zambelli}, G.~P. {Vacca} and R.~{Percacci},
  \emph{{Gravitational corrections to Yukawa systems}},
  \href{https://doi.org/10.1016/j.physletb.2010.04.043}{\emph{Physics Letters
  B} {\bfseries 689} (2010) 90}
  [\href{https://arxiv.org/abs/0904.0938}{{\ttfamily 0904.0938}}].

\bibitem{Eichhornetal1}
A.~{Eichhorn}, A.~{Held} and J.~M. {Pawlowski}, \emph{{Quantum-gravity effects
  on a Higgs-Yukawa model}},
  \href{https://doi.org/10.1103/PhysRevD.94.104027}{\emph{Physical Review D}
  {\bfseries 94} (2016) 104027}
  [\href{https://arxiv.org/abs/1604.02041}{{\ttfamily 1604.02041}}].

\bibitem{Eichhornetal3}
A.~{Eichhorn} and A.~{Held}, \emph{{Mass difference for charged quarks from
  quantum gravity}}, {\emph{ArXiv e-prints} (2018) }
  [\href{https://arxiv.org/abs/1803.04027}{{\ttfamily 1803.04027}}].

\bibitem{oda2016non}
K.~Oda and M.~Yamada, \emph{{Non-minimal coupling in Higgs-Yukawa model with
  asymptotically safe gravity}},
  \href{https://doi.org/10.1088/0264-9381/33/12/125011}{\emph{Classical and
  Quantum Gravity} {\bfseries 33} (2016) 125011}.

\bibitem{Eichhornetal2}
N.~{Christiansen}, A.~{Eichhorn} and A.~{Held}, \emph{{Is scale-invariance in
  gauge-Yukawa systems compatible with the graviton?}},
  \href{https://doi.org/10.1103/PhysRevD.96.084021}{\emph{Physical Review D}
  {\bfseries 96} (2017) 084021}
  [\href{https://arxiv.org/abs/1705.01858}{{\ttfamily 1705.01858}}].

\bibitem{RodigastSchuster}
A.~{Rodigast} and T.~{Schuster}, \emph{{Gravitational Corrections to Yukawa and
  ${\phi}^{4}$ Interactions}},
  \href{https://doi.org/10.1103/PhysRevLett.104.081301}{\emph{Physical Review
  Letters} {\bfseries 104} (2010) 081301}
  [\href{https://arxiv.org/abs/0908.2422}{{\ttfamily 0908.2422}}].

\bibitem{Martins2017PLB}
S.~{Gonzalez-Martin} and C.~P. {Martin}, \emph{{Do the gravitational
  corrections to the beta functions of the quartic and Yukawa couplings have an
  intrinsic physical meaning?}},
  \href{https://doi.org/10.1016/j.physletb.2017.09.011}{\emph{Physics Letters
  B} {\bfseries 773} (2017) 585}
  [\href{https://arxiv.org/abs/1707.06667}{{\ttfamily 1707.06667}}].

\bibitem{Martins2018JCAP}
S.~{Gonzalez-Martin} and C.~P. {Martin}, \emph{{Unimodular Gravity and General
  Relativity UV divergent contributions to the scattering of massive scalar
  particles}}, \href{https://doi.org/10.1088/1475-7516/2018/01/028}{\emph{JCAP}
  {\bfseries 1} (2018) 028} [\href{https://arxiv.org/abs/1711.08009}{{\ttfamily
  1711.08009}}].

\bibitem{Martins2018EPJC}
S.~{Gonzalez-Martin} and C.~P. {Martin}, \emph{{Scattering of fermions in the
  Yukawa theory coupled to unimodular gravity}},
  \href{https://doi.org/10.1140/epjc/s10052-018-5734-z}{\emph{European Physical
  Journal C} {\bfseries 78} (2018) 236}
  [\href{https://arxiv.org/abs/1802.03755}{{\ttfamily 1802.03755}}].

\bibitem{narain2017exorcising}
G.~{Narain}, \emph{{Exorcising Ghosts in Induced Gravity}},
  \href{https://doi.org/https://doi.org/10.1140/epjc/s10052-017-5249-z}{\emph{The
  European Physical Journal C} {\bfseries 77} (2017) 683}
  [\href{https://arxiv.org/abs/1612.04930}{{\ttfamily 1612.04930}}].

\bibitem{DeWittdynamical}
B.~S. DeWitt, \emph{Dynamical Theory of Groups and Fields}. Gordon and Breach,
  1965.

\bibitem{BunchParker}
T.~S. Bunch and L.~Parker, \emph{Feynman propagator in curved spacetime: A
  momentum space approach},
  \href{https://doi.org/10.1103/PhysRevD.20.2499}{\emph{Physical Review D}
  {\bfseries 20} (1979) 2499}.

\bibitem{tHooftandVeltman}
G.~`t~Hooft and M.~Veltman, \emph{Regularization and renormalization of gauge
  fields}, \href{https://doi.org/10.1016/0550-3213(72)90279-9}{\emph{Nuclear
  Physics B} {\bfseries 44} (1972) 189}.

\bibitem{BjorkenandDrell}
J.~D. Bjorken and S.~D. Drell, \emph{Relativistic {Q}uantum {F}ields}.
  McGraw-Hill, 1965.

\bibitem{BirrellandDavies}
N.~D. Birrell and P.~C.~W. Davies, \emph{Quantum fields in Curved Space}.
  Cambridge University Press, 1982.

\bibitem{FradkinTseytlin}
E.~S. Fradkin and A.~A. Tseytlin, \emph{On the new definition of the off-shell
  effective action}, {\emph{Nuclear Physics B} {\bfseries 234} (1984) 509}.

\bibitem{BarvinskyVilkovisky}
A.~O. Barvinsky and G.~A. Vilkovisky, \emph{The generalized
  {S}chwinger-{D}e{W}itt technique in gauge theories and quantum gravity},
  {\emph{Physics Reports} {\bfseries 119} (1985) 1}.

\bibitem{MossToms}
I.~G. Moss and D.~J. Toms, \emph{Invariants of the heat equation for
  non-minimal operators}, {\emph{Journal of Physics A: Mathematical and
  Theoretical} {\bfseries 47} (2014) 215401}.

\bibitem{Gilkey75}
P.~B. Gilkey, \emph{Recursion relations and the asymptotic behavior of the
  eigenvalues of the {L}aplacian}, {\emph{Compos. Math.} {\bfseries 38} (1979)
  201}.

\bibitem{Gilkey79}
P.~B. Gilkey, \emph{The spectral geometry of a {R}iemannian manifold},
  \href{https://doi.org/10.4310/jdg/1214433164}{\emph{J. Diff. Geom.}
  {\bfseries 10} (1975) 601}.

\bibitem{fulling1989aspects}
S.~A. Fulling, \emph{Aspects of Quantum Field Theory in Curved Spacetime}.
  Cambridge University Press, 1989.

\bibitem{avramidi2000heat}
I.~G. Avramidi, \emph{Heat Kernel and Quantum Gravity}, vol.~64. Springer,
  2000.

\bibitem{vassilevich2003heat}
D.~V. Vassilevich, \emph{Heat kernel expansion: user's manual},
  \href{https://doi.org/10.1016/j.physrep.2003.09.002}{\emph{Physics Reports}
  {\bfseries 388} (2003) 279}
  [\href{https://arxiv.org/abs/hep-th/0306138}{{\ttfamily hep-th/0306138}}].

\bibitem{kirsten2010spectral}
K.~Kirsten, \emph{Spectral Functions in Mathematics and Physics}. CRC Press,
  2010.

\bibitem{toms2014local}
D.~J. Toms, \emph{Local momentum space and the vector field}, {\emph{Physical
  Review D} {\bfseries 90} (2014) 044072}.

\bibitem{gusynin1997computation}
V.~P. Gusynin and V.~V. Kornyak, \emph{Computation of the
  {D}e{W}itt-{S}eeley-{G}ilkey coefficient $ e_4$ for nonminimal operator in
  curved space}, {\emph{Nuclear Instruments and Methods in Physics Research
  Section A: Accelerators, Spectrometers, Detectors and Associated Equipment}
  {\bfseries 389} (1997) 365}.

\bibitem{gusynin1999complete}
V.~P. Gusynin and V.~V. Kornyak, \emph{Complete computation of
  {D}e{W}itt-{S}eeley-{G}ilkey coefficient $ e_4$ for nonminimal operator on
  curved manifolds}, {\emph{arXiv preprint math/9909145} (1999) }.

\bibitem{tHooft1973}
G.~`t~Hooft, \emph{Dimensional regularization and the renormalization group},
  \href{https://doi.org/10.1016/0550-3213(73)90376-3}{\emph{Nuclear Physics B}
  {\bfseries 61} (1973) 455}.

\bibitem{Fujikawa1}
K.~Fujikawa, \emph{Path integral measure for gauge-invariant fermion theories},
  \href{https://doi.org/10.1103/PhysRevLett.42.1195}{\emph{Physical Review
  Letters} {\bfseries 42} (1979) 1195}.

\bibitem{Fujikawa2}
K.~Fujikawa, \emph{Comment on chiral and conformal anomalies},
  \href{https://doi.org/10.1103/PhysRevLett.44.1733}{\emph{Physical Review
  Letters} {\bfseries 44} (1980) 1733}.

\bibitem{Fujikawa3}
K.~Fujikawa, \emph{Path integral for gauge theories with fermions},
  \href{https://doi.org/10.1103/PhysRevD.21.2848}{\emph{Physical Review D}
  {\bfseries 21} (1980) 2848}.

\bibitem{Bonoraetal}
L.~{Bonora}, A.~D. {Pereira} and B.~L. {de Souza}, \emph{{Regularization of
  energy-momentum tensor correlators and parity-odd terms}},
  \href{https://doi.org/10.1007/JHEP06(2015)024}{\emph{Journal of High Energy
  Physics} {\bfseries 6} (2015) 24}
  [\href{https://arxiv.org/abs/1503.03326}{{\ttfamily 1503.03326}}].

\bibitem{TomsPRDscalar}
D.~J. Toms, \emph{Renormalization of interacting scalar fields in curved
  spacetime}, \href{https://doi.org/10.1103/PhysRevD.26.2713}{\emph{Physical
  Review D} {\bfseries 26} (1982) 2713}.

\end{thebibliography}
%\bibliographystyle{apsrev}

\end{document}